\documentstyle[12pt]{article}

\catcode`\@=11
\long\def\@makefntext#1{ %\parindent 1em
\protect\noindent \hbox to 3.2pt {\hskip-.9pt
$^{{\ninerm\@thefnmark}}$\hfil}#1\hfill} %can be used

 \def\@makefnmark{\hbox to 0pt{$^{\@thefnmark}$\hss}}  %original

\def\ps@myheadings{\let\@mkboth\@gobbletwo
\def\@oddhead{\hbox{} %\sl
\rightmark\hfil\ninerm\thepage}
\def\@oddfoot{}\def\@evenhead{\ninerm\thepage\hfil %\sl
\leftmark\hbox{}}\def\@evenfoot{}
\def\sectionmark##1{}\def\subsectionmark##1{}}

\begin{document}

%------------------------------------------------------------------------------
%MARCO FOR ABSTRACT BLOCK
\def\abstracts#1{{
	\centering{\begin{minipage}{30pc}\tenrm\baselineskip=12pt\noindent
	\centerline{\tenrm ABSTRACT}\vspace{0.3cm}
	\parindent=0pt #1
	\end{minipage} }\par}}

%------------------------------------------------------------------------------
%NEW MACRO FOR BIBLIOGRAPHY
\newcommand{\bibit}{\it}
\newcommand{\bibbf}{\bf}
\renewenvironment{thebibliography}[1]
	{\begin{list}{\arabic{enumi}.}
	{\usecounter{enumi}\setlength{\parsep}{0pt}
%1.25cm IS STRICTLY FOR PROCSLA.TEX ONLY
\setlength{\leftmargin 1.25cm}{\rightmargin 0pt}
%0.52cm IS FOR NEW DATA FILES
%\setlength{\leftmargin 0.52cm}{\rightmargin 0pt}
	 \setlength{\itemsep}{0pt} \settowidth
	{\labelwidth}{#1.}\sloppy}}{\end{list}}

%------------------------------------------------------------------------------
%FOLLOWING THREE COMMANDS ARE FOR 'LIST' COMMAND.
\topsep=0in\parsep=0in\itemsep=0in
\parindent=1.5pc

%LIST ENVIRONMENTS
\newcounter{itemlistc}
\newcounter{romanlistc}
\newcounter{alphlistc}
\newcounter{arabiclistc}
\newenvironment{itemlist}
    	{\setcounter{itemlistc}{0}
	 \begin{list}{$\bullet$}
	{\usecounter{itemlistc}
	 \setlength{\parsep}{0pt}
	 \setlength{\itemsep}{0pt}}}{\end{list}}

\newenvironment{romanlist}
	{\setcounter{romanlistc}{0}
	 \begin{list}{$($\roman{romanlistc}$)$}
	{\usecounter{romanlistc}
	 \setlength{\parsep}{0pt}
	 \setlength{\itemsep}{0pt}}}{\end{list}}

\newenvironment{alphlist}
	{\setcounter{alphlistc}{0}
	 \begin{list}{$($\alph{alphlistc}$)$}
	{\usecounter{alphlistc}
	 \setlength{\parsep}{0pt}
	 \setlength{\itemsep}{0pt}}}{\end{list}}

\newenvironment{arabiclist}
	{\setcounter{arabiclistc}{0}
	 \begin{list}{\arabic{arabiclistc}}
	{\usecounter{arabiclistc}
	 \setlength{\parsep}{0pt}
	 \setlength{\itemsep}{0pt}}}{\end{list}}

%------------------------------------------------------------------------------
%ACKNOWLEDGEMENT: this portion is from John Hershberger
\def\@citex[#1]#2{\if@filesw\immediate\write\@auxout
	{\string\citation{#2}}\fi
\def\@citea{}\@cite{\@for\@citeb:=#2\do
	{\@citea\def\@citea{,}\@ifundefined
	{b@\@citeb}{{\bf ?}\@warning
	{Citation `\@citeb' on page \thepage \space undefined}}
	{\csname b@\@citeb\endcsname}}}{#1}}

\newif\if@cghi
\def\cite{\@cghitrue\@ifnextchar [{\@tempswatrue
	\@citex}{\@tempswafalse\@citex[]}}
\def\citelow{\@cghifalse\@ifnextchar [{\@tempswatrue
	\@citex}{\@tempswafalse\@citex[]}}
\def\@cite#1#2{{$\null^{#1}$\if@tempswa\typeout
	{IJCGA warning: optional citation argument
	ignored: `#2'} \fi}}
\newcommand{\citeup}{\cite}
\newcommand{\emet}{{\em et al.}}
\newcommand{\csb } {charge symmetry breaking}
%------------------------------------------------------------------------------
\font\twelvebf=cmbx10 scaled\magstep 1
\font\twelverm=cmr10 scaled\magstep 1
\font\twelveit=cmti10 scaled\magstep 1
\font\elevenbfit=cmbxti10 scaled\magstephalf
\font\elevenbf=cmbx10 scaled\magstephalf
\font\elevenrm=cmr10 scaled\magstephalf
\font\elevenit=cmti10 scaled\magstephalf
\font\bfit=cmbxti10
\font\tenbf=cmbx10
\font\tenrm=cmr10
\font\tenit=cmti10
\font\ninebf=cmbx9
\font\ninerm=cmr9
\font\nineit=cmti9
\font\eightbf=cmbx8
\font\eightrm=cmr8
\font\eightit=cmti8

%----------------------START OF DATA FILE------------------------------
%\pagestyle{empty}
\overfullrule=0pt
%------------------------------------------------------------------------------
%MARCO FOR ABSTRACT BLOCK
\def\abstracts#1{{
	\centering{\protect\begin{minipage}{30pc}\small\baselineskip=12pt\noindent
	\centerline{\small ABSTRACT}\vspace{0.3cm}
	\parindent=0pt #1
	\end{minipage} }\par}}

%------------------------------------------------------------------------------
\begin{center}

{CHARGE INDEPENDENCE AND CHARGE SYMMETRY  }
%of Nuclear Forces}

\vspace{.5in}

GERALD A. MILLER \\

{\em Department of Physics, FM-15, University of Washington, Seattle, WA 98195}

\vspace{.25in}

and \\
\vspace{.25in}
WILLEM T.H. VAN OERS \\
{\em Department of Physics, University of Manitoba,
Winnipeg, MB,  R3T 2N2, Canada \\
and \\
TRIUMF, 4004 Wesbrook Mall, Vancouver, B.C., V6T 2A3, Canada}
\end{center}
\vspace{.25in}

\abstracts%
{Charge independence and charge symmetry are approximate symmetries of nature,
violated by the perturbing effects of the mass difference between up and  down
quarks and by electromagnetic interactions. The observations of the symmetry
breaking effects in nuclear and particle  physics and the implications of
those effects are reviewed.}
\vspace{.25in}

\noindent{\bf 1. Introduction}

\vspace{0.3cm}

This paper is concerned with  charge  independence and charge symmetry which
provide powerful tools in organizing and describing  the multiplet structure of
hadrons and nuclei\cite{pttext,nptex}.  These symmetries are imperfect;
diverse  small but interesting  violations have been discovered. For reviews
see Refs. [3]-[7].

The emergence of Quantum Chromodynamics, QCD, as the underlying theory of the
strong interaction has added a new impetus to this field. QCD indicates that
each and every violation of charge independence or charge symmetry has its
origin in the different masses of the up and down current quarks and in the
electromagnetic interactions between the quarks.  Thus the small imperfections
of these symmetries provide a unique  opportunity to study the relation between
quark mass differences and  hadronic and nuclear observables. A prominent
example is that the positive value of $m_d-m_u$ causes the neutron to be
heavier than the proton in  contrast with the result expected from
electromagnetic effects. This review is concerned with the evidence that  the
light quark mass difference $m_d$-$m_u$ plus electromagnetic effects accounts
for charge independence and charge symmetry breaking, CSB, in systems of baryon
number ranging from 0 to 208.

Let us define the terms charge independence and charge symmetry. Consider the
QCD Lagrangian  in the limit that $m_d$ and $m_u$ vanish and ignore
%%% change of wording
electromagnetic effects. In that case,  the $u$ and $d$ quarks are equivalent
and can  be treated as an isodoublet $\left({u\atop d}\right)$.  One may
introduce the isospin operators  $\vec\tau$ with
\begin{eqnarray}
[\tau_i,\tau_j] = i\;\epsilon_{ijk}\tau_k,
\end{eqnarray}
\begin{eqnarray}
\tau_3|u> = |u>, \qquad
\tau_3|d> = -|d>.  \label{e:iso}
\end{eqnarray}
The total isospin for a system of quarks is then
\begin{eqnarray}
\vec T =\Sigma_i\vec\tau(i)/2.
\end{eqnarray}
  In the present limit,
\begin{eqnarray}
[H,\vec T] = 0.
\end{eqnarray}
This vanishing is the invariance under any rotation in isospin space and known
as charge independence or  isospin symmetry. Charge symmetry is related
specifically
to a rotation
by $\pi$ about the y-axis in isospin space:
\begin{eqnarray}
[H,P_{cs}]=0,
\end{eqnarray}
with
\begin{eqnarray}
P_{cs} = e^{i\pi T_2},
\end{eqnarray}
if the positive z-direction is associated with positive charge.
$P_{cs}$ converts $u$ quarks into $d$ quarks and vice versa:
\begin{eqnarray}
P_{cs}|u>
= -|d>, \quad P_{cs}|d> = |u>. \label{e:pcs}
\end{eqnarray}
Ref.~[6] was the first to use this  quark based definition of
charge symmetry.

A  violation of charge symmetry implies that charge  independence is broken;
but the converse is not true. For example, the operator $\tau_3(1)\tau_3(2)$
between two nucleons preserves charge symmetry.

It is hadrons, not quarks,  that are observed, so it is reasonable to ask how
is the quark operator $\vec T$ related to hadronic isospin operators.  Isospin,
like charge,  is an additive quantum number, so the  isospin operator can
be expressed equivalently  in terms of quarks  or hadrons. This means that the
hadronic implications of charge independence or  charge symmetry can be
understood from the quark content of the systems  involved. For example, charge
symmetry predicts the equality of the $\Lambda (uds) - p(uud)$ and $\Lambda
(uds) - n(udd)$ interactions.

But there  are other serious  concerns about the application of these quark
concepts to reality.   The first problem is that quarks are confined and
can not be isolated. This  means that
a  specific definition of the term ``quark mass" is required. It is natural to
speak of the mass that appears in the Lagrangian of QCD; this is the  current
quark mass. Evaluations of these masses  have been made by many authors.  For
example, Gasser and Leutwyler\cite{GL82} use QCD sum rules for the divergence
of the  axial vector current along with current  algebra and pseudoscalar meson
masses to obtain $m_u=5.1\pm 1.5$~MeV and $m_d=8.9\pm 2.6$~MeV (evaluated  at
a momentum transfer scale of 1~GeV). The individual quark masses have an  error
of $\pm 30\%$, while the ratio $m_d/m_u$ is more accurately  determined as 1.76
$\pm$ 0.13 \cite{GL82}. See also Ref.~[9]. %%
The scale dependence arises  because the bare masses
are replaced by physical ones by the necessary  renormalization procedure. But
the physical masses are not measured, so that it is necessary to choose an
arbitrary kinematic reference point to define the physical quark masses.

A second problem is that  the ratio ${m_d\over m_u}\approx 2$ seems to strongly
violate the symmetry. Thus one may wonder why any trace of charge independence
would remain in nature. However each of $m_d$ and $m_u$ is less than 10~MeV,
and these current quark masses are replaced by constituent
masses for confined quarks.
In this case, the mass of each of the three quarks  in the nucleon
supplies about one-third of the nucleon mass, which is much larger than the
current quark mass. Various strong interaction effects, every one of which
respects charge independence, are thought to cause this relatively large
value. Thus the ratio of the down to up constituent quark mass is very close
to unity despite  the fact that the ratio of the current quark masses is about
two! Explanations of the origin of this unity fall into  two categories. The
first is dynamical symmetry breaking,  in which non-perturbative
quark-pair creation and gluon-exchange effects cause the vacuum to reduce its
energy by acquiring a non-zero expectation value (vacuum condensate)  of the
quark field  operator $\overline{\psi}(0)\psi(0)$.  In that case, the color
force must lead also to a quark self-energy  that plays the role of a quark
mass. Numerical calculations starting with  the chiral limit ($m_d=0, m_u=0$)
show that dynamical symmetry breaking generates quark masses consistent
with the large constituent
quark masses\cite{dsb,dsbcsb}.  In the chiral limit, the resulting constituent
masses of the up and down quarks are the same. One can also include the small
current quark masses with the result\cite{dsb,dsbcsb} that  the small nature of
the mass difference between the up and down current quarks is retained  by the
resulting constituent quarks, and is small compared with the sum of the
constituent masses.

In constituent quark models
the effects of confinement can also yield contributions to the
quark mass because the separation
between a confining potential and a quark mass is
a matter of choice.
Other models are that of the MIT bag and the non-topological soliton
with massless or very light quarks. Then   the average quark
kinetic energy (of order 350~MeV) acts as a quark mass in many matrix elements.
In any case,  the confining potential and kinetic energy of massless quarks are
the same for up and down quarks. The average constituent
quark mass, whether arising from dynamical symmetry breaking
or confinement, is dominated by charge
symmetric effects and is about a third of the proton mass. In all models of QCD
that involve quarks, any small  %%
difference between up and down quark constituent quark
masses is caused by  a difference between the corresponding current quark
masses.

A final comment about quarks concerns the term ``accidental symmetry",
introduced by Gross, Treiman and Wilczek\cite{GTW}, to describe isospin
symmetry. The term ``accidental" arises  because charge independence is
well-maintained in nature even  though  the current quark %%
ratio $m_d/m_u$ is quite large.
But isospin symmetry naturally results from the dependence of  observables on
constituent quark masses $m({\rm const})$ which satisfy  $m_d({\rm
const})/m_u({\rm const}) \approx 1$ as an inevitable  consequence of dynamical
symmetry breaking inherent in QCD. Furthermore, van Kolck\cite{VK93} has used
the chiral symmetry of QCD to show that  charge  independence breaking  effects
are small;  of the order of $(m_d-m_u)/m_\rho$  where $m_\rho \approx 780 \,
MeV$ is the mass of the rho meson. (A possible  %%
exception is $\pi^0$-nucleon scattering for which charge dependence {\bf is}
governed by the ratio $(m_d-m_u)/(m_d+m_u)$ of
current quark masses\cite{VK93}.)
Thus we believe that progress in
understanding non-perturbative aspects of QCD is now sufficient to say that the
use of the term accidental symmetry is a misnomer. The approximate validity of
charge independence is understood.

The remainder of this article is concerned first with the quark mechanisms
behind charge independence and its breaking, Sect.~2. The mass difference
between the up and down quarks leaves its imprint on many hadronic  matrix
elements. This allows the computation of many nuclear effects that  break
charge independence. Sect.~3. is concerned with the diverse sets of  new
experimental  information. Several significant observations have been made, but
the  interpretation of some  other experiments is  clouded by the presence of
Coulomb effects. A summary and discussion of the directions that future work
could take  is given in Sect.~4.

\vspace{0.5cm}
\noindent
{\bf 2. Mechanisms of charge independence and charge symmetry breaking}

\vspace{0.3cm}
The breaking of charge independence and charge symmetry occurs via the
positive value of $m_d-m_u$ and, by quark electromagnetic  effects. How are
these effects  realized in nature?  We discuss the present ideas and argue
that, at present, the most successful procedure is to use quark effects to
understand hadronic masses  %%
and meson-mixing. Then  nucleon-nucleon
interactions are described in terms of meson  exchange models which include the
effects of hadronic charge independence breaking mandated by the quark effects.

\vspace{0.4cm}
\noindent
{\em 2.1 Mass splittings of hadronic isospin multiplets}

\vspace{0.2cm}
There are many quark models of baryon and meson  structure\cite{pttext}. In
those models one simply evaluates the  consequences of the quark masses and of
the electromagnetic interactions. A common explanation of the mass differences
within isotopic multiplets  has emerged. This is important since an old and
difficult issue --- ``the sign of the isotopic mass splitting puzzle" was
resolved by the use  of quarks. Before quark physics was introduced, the only
definite mechanism  for the breaking of charge independence was the
electromagnetic interaction. Charge independence implies that the neutron and
proton mass are equal. But naive estimates and sophisticated
evaluations\cite{C63} based on the  electromagnetic interaction gave the
inevitable result that a charged particle is heavier than its neutral partner.
But nature does not follow this recipe: $m_n>m_p$ and $m_{K^0}>m_{K^-}$, while
$m_{\pi^0}< m_{\pi^+}$; furthermore $m_{\Sigma^+}<m_{\Sigma^0}<m_{\Sigma^-}$.
This riddle is solved by  realizing that we should not arrange multiplets by
their electric charge  but by their u-d flavor. If two hadrons are related by
replacing a u quark by a d quark,  the  d-rich system is heavier. This holds
for  all mesons and baryons\footnote{The $\Sigma_c^0(ddc)$ may be slightly
less
massive than the $\Sigma_c^+(udc)$, but the  uncertainty in the $\Sigma_c^+$
mass is $\pm 3.1$ MeV\cite{pdg} which is large in the present context.} for
which a comparison is possible. See   Table~1 of Ref.~[7]. Thus the
positive value of $m_d-m_u$ is more important numerically  than the
electromagnetic effects.

The above discussion is qualitative, but the details of many complicated
computations have been reviewed in Ref.~[6]. The positive nature of
$m_d-m_u$, as corrected  by electromagnetic effects,  accounts for the mass
differences within hadronic multiplets even though the specific values vary
from model to model.

\vspace{0.4cm}
\noindent
{\em 2.2 Mixing Hadrons}

\vspace{0.2cm}
In the absence of charge independence breaking the neutral mesons of u-d
flavor are  states of pure isospin, given schematically as
\begin{eqnarray}
|I=1>={1\over \sqrt 2}|u\bar u> -{1\over \sqrt 2}|d\bar d>,\qquad
|I=0>={1\over \sqrt 2}|u\bar u> +{1\over \sqrt 2}|d\bar d>.\qquad
\end{eqnarray}
The isospin of a state is determined by the final states obtained via  strong
decay processes, i.e. 2$\pi$ for I=1 and 3$\pi$ for I=0. However, the
perturbing effects of the  quark mass difference and electromagnetic effects
cause the states to mix. For example, the quark mass contribution to the  QCD
Hamiltonian, $H_m$
\begin{eqnarray}
H_m=m_d \bar{d} d +m_u \bar{u} u,
\end{eqnarray}
gives a mixing matrix element of the form
\begin{eqnarray}
<I=1|H_m|I=0>=m_u-m_d,
\end{eqnarray}
which is negative.  Electromagnetic effects also enter, as we shall discuss.

In general, neutral mesons are mixtures of I=0 and I=1 states,
with the largest mixing occurring when the unperturbed states are close in
mass. Thus the best studied case is that of $\rho^0-\omega$ mixing.
The case of $\pi^0-\eta-\eta^\prime$ mixing is handled in an analogous
fashion, but the octet-singlet SU(3) mixing between the $\eta $ and the
$\eta^\prime$ causes a complication. This is reviewed in Sect. 3.3.12 of
Ref.~[6].

\begin{figure}
  \vspace{3.5in}
  \caption{Amplitudes for $e^+e^-\to\pi^+\pi^-$.}
\end{figure}

Here we concentrate on $\rho^0-\omega$ mixing which is  the strongest and most
prominent observation  of charge symmetry breaking. The effects of this matrix
element are observed\cite{Q78,B85} in the annihilation process
$e^+e^-\to\pi^+\pi^-$.  The relevant diagrams are shown in Fig.~1 and the huge
signal arising from the small widths of the $\omega$-meson is  displayed in
Fig.~2.  The mixing matrix element has been extracted\cite{CB87}  to be
\begin{equation}
<\rho^0|H|\omega > = -4520 \pm 60~{\rm MeV}^2 .
\end{equation}
This matrix element is expressed in units of mass$^2$, because the
observable is related to the self-energy that appears in the Klein-Gordon
equation governing meson propagation, so the meson states are
normalized as
$<\vec p,\alpha|\vec p \, ',\alpha> = 2 (\vec p\cdot \vec p \, ' +m_\alpha^2)
(2\pi)^3 \delta (\vec p -\vec p \, ')$. If one removes the momentum-conserving
delta function from the normalization, the states can be normalized to unity
e.g. $<\rho\| 1 \|\rho>=1$. In this basis,
 the mixing matrix element can be expressed as
$<\rho^0\|H\|\omega >
=<\rho^0|H|\omega >/(m_\rho + m_\omega)\approx 2.9 $ MeV.

\begin{figure}
  \vspace{3.4in}
  \caption{$\sigma(e^+e^-\to \pi^+\pi^-$).  These are the data introduced
and summarized in Ref.~[16].}
\end{figure}

The  extracted
matrix element $<\rho^0|H|\omega >$
includes the effect of the electromagnetic process depicted
in Fig.~3. The  quantities $f_\rho$ and $f_\omega$ have been determined from
the processes $e^+e^-\to\rho,\omega\to e^+e^-$.   The most recent
analysis\cite{L79}  gives $<\rho^0|H_{em}|\omega> = 640\pm 140$ MeV$^2$ so that
the strong contribution $(H=H_{str} + H_{em}$) is given by
$<\rho^0|H_{str}|\omega> = -5160 \pm 150 $ MeV$^2$.  Another notable feature is
that the electromagnetic contribution to the $\rho\omega$-mixing self-energy is
of the form
\begin{equation}
\Pi^{em}_{\rho\omega}(q^2)\sim {e^2\over f_\rho f_\omega}\;{1\over q^2}
\end{equation}
where $q^2$ is the square of the vector meson four-momentum.

\pagebreak
The question of whether the consequences of this significant mixing has any
observable consequences in nucleon-nucleon scattering and nuclear physics is
discussed at the end of the Sect.~2.3.
\begin{figure}
  \vspace{0.5in}
  \caption{Electromagnetic contribution to $\rho^0$-$\omega$ mixing.}
\end{figure}

Baryons can also mix. The oldest example\cite{DV64,GS67} is that of
$\Lambda-\Sigma^0$ mixing. See the review, Ref.~[22]. In first-order
perturbation theory
\begin{eqnarray}
|\Lambda> & = & |uds, I=0>+\alpha|uds, I=1> \\
|\Sigma^0> & = & -\alpha |uds, I=0> +|uds, I=1>,
\end{eqnarray}
where
\begin{eqnarray}
\alpha={<uds, I= 1 |H|uds, I=0>\over (M_\Lambda-M_{\Sigma^0})}.
\end{eqnarray}
\noindent
The use of SU(3) symmetry\cite{DV64} gives $\alpha=-0.013$, and the quark model
of Ref.~[21] gives $\alpha=-0.011$, so the mixing  matrix element is
about 1~MeV. The charge dependent operator must connect the (ud+du) of the
$\Sigma^0$  with the (ud-du) of the $\Lambda$, so  the quark mass and  kinetic
energy difference, and Coulomb  interaction do not contribute. The one gluon
exchange interaction which  causes the $\Lambda-\Sigma^0$ splitting  has a
charge dependence arising from  its dependence on the quark masses. An order of
magnitude estimate then is  that $|\alpha|\approx {m_d-m_u\over \bar m}\approx
3.3 MeV/340 \, MeV\approx 0.01$
where the numerator is from Ref.~[7] and
the denominator is  a typical constituent quark mass.

Karl\cite{GK94} and Henley \&
Miller\cite{HM94} have studied the influence of  $\Lambda-\Sigma^0$
mixing on the beta decays of $\Sigma^\pm$ to the  $\Lambda$.
The decay rates are known\cite{pdg} to about 5\% for the $\Sigma^-$ but to
about 25 \% for the $\Sigma^+$, with about a two standard deviation
difference  between central values. However, such a difference arises from
effects of the $\Sigma^\pm$ mass differences
on the phase space. Phase space effects can be calculated precisely and
are not interesting. However  the effect of
$\Lambda-\Sigma^0$ mixing is predicted to cause a substantial
$\approx $ 6\% difference between the squares of the axial vector matrix
elements. Thus Karl and Henley \& Miller
suggest  that
improved experiments at about the 1\% level in the ratio of the decay rates
could clearly observe the effects of baryon mixing.
A similar mixing is expected for the charmed, bottom, and top versions of the
$\Lambda$ and $\Sigma$ baryons with charge symmetry breaking effects
exhibited in their beta decays.

\vspace{0.3cm}
\noindent
{\em 2.3  Two nucleon scattering:  quark models and meson exchange models}

\vspace{0.2cm}
There have been a number of calculations using non-relativistic quark and  bag
models to compute nucleon-nucleon scattering\cite {QUARKCALCS} and some of
these have been applied to compute charge independence breaking  of the
nucleon-nucleon scattering lengths. The main motivation for using the quark
model is to gain a better understanding of the nucleon-nucleon  short-ranged
repulsion. This, in  present  quark models, arises from the quark Pauli
principle and from the gluon exchange hyperfine  interaction.  Then the charge
independence breaking can be computed  directly in terms of  the mass
difference between composite quarks and electromagnetic effects.

Some of the  quark models that are most successful in reproducing observed
phase shifts use the  resonating group model (RGM)  in which  the six-quark
wave function is  an antisymmetrized product of six single quark wave
functions. This approach has been criticized by Miller\cite{M89} who argued
that the inclusion of gluon degrees of freedom cause the exchange terms
introduced by the antisymmetrizing operator to vanish.

Nonetheless, quark models can be used to provide reasonable descriptions of the
nucleon-nucleon phase shifts if the long-range effects of $\pi$  and $\sigma$
exchange between nucleons are also included.
Brauer, Faessler and Henley \cite{BFH85} used the RGM to compute
the strong interaction contribution to the difference between the nn and pp
scattering lengths,
$a_{nn}^N$-$a^N_{pp}$. Charge symmetry breaking
contributions from the one-gluon (magnetic) and quark kinetic energy were
included. These two mechanisms tend to have cancelling effects, so that the
resulting charge symmetry breaking
is small: $a_{pp}^N$-$a_{nn}^N$ = 0.5~fm. See also Wang, Wang and
Wong\cite{WWW88} who find that the influence of  quark effects on charge
independence breaking  are highly model dependent.  Six-quark bag models have
also been used to compute charge independence breaking effects.  The results of
Koch and Miller\cite{KM85} for $a_{pp}^N$-$a_{nn}^N$ are similar to those of
Ref.~[27], but depend strongly on the model parameters.

The principal finding is that the quark-model effects responsible for the
``hard" core of the nucleon-nucleon interaction  harbor only small charge
independence breaking effects. It is difficult to precisely say how small  the
effects are because of the strong model dependence of current  treatments.
Instead it is natural to look for  charge independence and charge symmetry
breaking arising from effects of  longer range. Thus the remainder of our
discussion employs meson exchange models.

There has been much discussion of how meson exchange leads to the breaking  of
charge independence; see the reviews [4]-[6]. The longest range
force arises from the one pion exchange potential (OPEP), which also supplies
significant breaking of charge independence.  This is due to the relatively
large mass difference. ${m_{\pi^\pm} - m_{\pi^0}\over m_{\pi^0}}\approx 0.04$.
One might worry about including the charge dependence of the coupling constants
for neutral ($g_0$) and charged ($g_c$) pions. However $g^2_0 = g^2_c$ to
better than about 1\%, according to recent phase shift analyses of Bugg and
Machleidt\cite{BM94} and the Nijmegen group\cite{NIM}. The quark model
predictions for the  charge  dependence of the coupling constants are reviewed
in Ref.~[6]; there is no consensus on the results.

One must also include the effects of the $\pi$ mass difference in the two pion
exchange potential (TPEP).  Henley and Morrison\cite{HM66} were the first to do
that. Similarly one may include the effects of isotopic mass differences of
heavier  mesons and  the baryons which appear in intermediate states.
Those effects are substantially smaller than the one already mentioned.

Another mechanism involves the two--boson exchange force that arises when  one
of the bosons is a photon. The pion is the lightest meson so, of these,
the  $\pi\gamma$
exchange leads to  the longest range force and was expected to be the  most
important. Early calculations were done in the static limit, and later
calculations\cite{C75,B75} are not in the refereed literature or have been
criticized  in Ref.~[5].

It is natural to study the
mechanisms mentioned here by predicting nucleon-nucleon scattering
observables\cite{LS80}.
The information available is limited mainly to the $^1$S$_0$
channel\cite{MNS90}, although
there are some new results.
The  Nijmegen group\cite{KKR94} has obtained a multi-energy partial wave
analysis of all nucleon-nucleon scattering data for laboratory kinetic energies
below  350 MeV. An accurate determination of all phase shifts and mixing
parameters has been obtained.
A charge dependent and charge asymmetric nucleon-nucleon potential has been
recently fit to these data by Wiringa et al\cite{WSS94}.
The charge independence breaking arises from
the effect of the pion mass differences and from phenomenological
modifications of the shorter range forces required to reproduce the data.
The charge asymmetry is
obtained by allowing the strengths of the central
S=0, T=1 nn and pp forces to differ in a way that agrees with the low
energy nucleon-nucleon scattering data.
A different  technique is to use
the nucleon-deuteron
break up reactions which allows the determination
of a set
of nn, np and pp $^3$P$_j$ phase shifts\cite{GW93}. This does not yield
a unique
determination of these phases\cite{Gloe}.

\vspace{0.3cm}
\noindent
a) $^1$S$_0$ scattering lengths

\vspace{0.2cm}

We  discuss
charge independence breaking of
the nucleon-nucleon  scattering lengths
in the  $^1$S$_0$ channel to illustrate how the
various mechanisms mentioned above really work.

Charge independence, $[H,\vec T]=0$, imposes the equalities of the
nucleon-nucleon scattering lengths $a_{pp} = a_{nn} = a_{np}$.  But
electromagnetic effects are large and it is necessary to make corrections.
The results are analyzed, tabulated and discussed in Ref.~[6] and
updated here in Sect.~3.1; see Table~3.   These are
\begin{eqnarray}
a_{pp}^N & = &-17.3\pm 0.4\;{\rm fm}\nonumber \\
a_{nn}^N & = & -18.8\pm 0.3\;{\rm fm} \\
a_{np}^N & = & -23.75\pm 0.09\;{\rm fm}, \nonumber
\end{eqnarray}
in which the superscript $N$ represents the ``nuclear" effect obtained  after
the electromagnetic corrections have been made. The differences between these
scattering lengths represent charge  independence and charge symmetry breaking
effects.  There are very large percentage differences between these numbers
which may seem surprising.  But one must recall that that is the inverse of the
scattering lengths that are related to the potentials.  For two different
potentials, $V_1,V_2$ the scattering lengths $a_1$, $a_2$ are related by
\begin{equation}
{1\over a_1} - {1\over a_2} = M\int^\infty_0 dr\;u_1(V_1-V_2)u_2
\label{e:delta}
\end{equation}
where $u_1$ and $u_2$ are the wave functions, normalized so that their limit
at large r is  1-r/a$_i$ and u(0)=0.
Evaluating Eq.~(\ref{e:delta}) leads to the result\cite{H69},
\begin{equation}
{\Delta a\over a} = (10-15){\Delta V\over V} ,\label{e:deltaa}
\end{equation}
where the variation between 10 and 15 arises from using different radial
shapes for V(r).
One defines $\Delta a_{CD}$ to measure the charge independence breaking, with
\begin{equation}
\Delta a_{CD} \equiv
{1\over 2} (a_{pp}^N + a_{nn}^N) - a_{np}^N = 5.7\pm 0.3\;{\rm fm} .
\end{equation}
The charge dependence breaking is then about  about 2.5\% if one uses
Eq.~(\ref{e:deltaa}).  The breaking of charge symmetry CSB is represented
by the quantity
\begin{equation}
\Delta a_{CSB} \equiv a_{pp}^N - a_{nn}^N = 1.5\pm 0.5\;{\rm fm} .
\end{equation}

Some computations\cite{HM66,EM83,CM86}  of $\Delta a_{CD}$ are displayed in
Table~1.  The agreement with the experimental value of $\Delta
a_{CD} = 5.7\pm 0.5$ fm is very good.  The errors allow some room
for other effects, including those due to explicit
quark effects.  It is clear that the understanding of charge
dependence has been rather good.

\vspace{0.5cm}

\centerline{Table~1. Calculations of $\Delta a_{CD}$ (fm)}

\vspace{0.2in}
\begin{center}
\begin{tabular}{|c|c|c|c|}
\hline
& Henley, Morrison \cite{HM66}&Ericson, Miller\cite{EM83}&Cheung, Machleidt
\cite{CM86} \\
& 1966 & 1983 & 1986 \\ \hline
OPEP & 3.5 & 3.5 $\pm$ 0.2 & 3.8 $\pm$ 0.2$^a$ \\
TPEP (all) & 0.90 & 0.88 $\pm$ 0.1 & 0.8 $\pm$ 0.1 \\
Coupling & {\small b} & 0$^c$ & \\
Constants &&& \\
$\gamma\pi$ & & 1.1 $\pm$ 0.4$^d$ & 1.1 $\pm$ 0.4$^d$ \\ \hline
Total & & 5.5 $\pm$ 0.3 & 5.7 $\pm$ 0.5 \\ \hline
\end{tabular}
\end{center}
\vspace{0.2cm}
\noindent
a. This also includes the effects of $\pi\rho,\pi\omega$ and
$\pi\sigma$ exchanges. \\
\noindent
b. Henley and Morrison showed that one could choose charge dependent coupling
constants to describe the remainder of $\Delta a_{CD}$, but these were
unknown. \\
\noindent
c. The effect of using charge dependent coupling constants tends to
cancel if these are used consistently in OPEP and TPEP. \\
\noindent
d. This is an average\cite{EM83} of the results of Refs.~[33]
and [34].

\vspace{0.5cm}

Now consider  the charge symmetry breaking mechanisms
responsible for  the non-zero value of  $\Delta
a_{CSB}$. The previously dominant one-pion exchange effect is absent here since
only neutral pions are exchanged.  Other mechanisms are needed, with  the
exchange of a mixed $\rho^0\omega$ meson  a natural choice. This is shown in
Figs.~4a and 4b.  The electromagnetic contribution, Fig.~4b, is part of the
long
range, electromagnetic interaction as modified by the vector meson contribution
to  the form factors.  The strong interaction term gives a nucleon-nucleon
force of a medium range.  This leads to a contribution to $\Delta a_{CSB}$ of
1.4 fm, obtained by rescaling the Coon and Barrett\cite{CB87} result by the
ratio $1.1\left({5160\over 4652}\right)$.  This accounts for the observed
effect $\Delta a_{CSB} = 1.5 \pm 0.5$ fm, while other  effects seem
small\cite{CB87}.
\begin{figure}
  \vspace{2.0in}
  \caption{$\rho^0$-$\omega$ exchange contributions (a) Short range, strong
        interaction effect; (b) Long range, electromagnetic effect}
\end{figure}

This agreement with experiment may not be satisfactory.  A significant
extrapolation is involved since $<\rho^0|H_{str}|\omega>$ is determined at $q^2
= m^2_\rho$, while in the NN force the relevant $q^2$ are spacelike, less than
or equal to zero.  Goldman, Henderson and Thomas\cite{GHT92} investigated the
possible $q^2$  dependence of $<\rho^0|H_{str}|\omega>$ by evaluating the
diagram  of Fig.~5 using free quark propagators.  They obtained a substantial
$q^2$ dependence.  The use of such a $<\rho^0|H_{str}|\omega>$ obliterates  the
resulting charge symmetry breaking
potential and its effects in nucleon-nucleon scattering\cite{IN94}.
Similar results for the mixing matrix element were also
obtained in the work of Refs.~[44]-[48]. Furthermore,
O'Connel at al.\cite{OCPTW94} have argued that within a broad class of models
the amplitude for $\rho^0$-$\omega$ mixing must vanish at the transition from
timelike to spacelike four momentum. However, this result is obtained by
assuming that ``there is no explicit mass-mixing term in the bare Lagrangian."
QCD has such a term and we therefore  expect that any equivalent
effective
hadronic theory would contain such a term. Thus, this vanishing need not occur.

\begin{figure}
  \vspace{0.7in}
  \caption{Quark model of $\rho^0$-$\omega$ mixing.}
\end{figure}

Our view is that the charge symmetry breaking effects of the $d$-$u$ mass
difference in vector exchanges must persist, with little variation in $q^2$,
whether one works directly with quarks or one uses hadronic matrix
elements to capture the quark effects.
However, we examine the consequences of the idea that
$<\rho^0|H_{str}|\omega>$ does have a strong variation with $q^2$.
\begin{figure}
  \vspace{3.0in}
  \caption{Model of Krein, Thomas and Williams -
$q^2$ variation of $<\rho^0|H_{str}|
\omega>$}
\end{figure}

Consider the results of the ``minimal" model of Krein, Thomas and
Williams\cite{KTW93} which are displayed in Fig.~6.  This work models
confinement in terms of pole-less quark propagators.  Here we emphasize
Miller's  argument\cite{M94} that  models which obtain the $q^2$ dependence of
$<\rho^0|H_{str}|\omega>$ from the diagram of Fig.~4 have an implicit
prediction for the $q^2$ variation of the $\rho$-$\gamma^*$ transition matrix
element $e/f_\rho(q^2)$, see Fig.~7.  Miller's  evaluation of this using the
minimal model of Ref.~[45] is shown in Fig.~8.  A significant variation
is seen, with a gain of a factor of four in the magnitude of $e/f_\rho (q^2)$
as $q^2$ is increased from 0 to $m_\rho^2$.
This is a noteworthy observation because $f_\rho (q^2)$ can be extracted from
$e^+e^-\to\rho\to e^+e^-$ data at $q^2 = m^2_\rho$ and from the high energy
$\gamma p\to\rho^0 p$ reaction at $q^2 = 0$.  The results of many experiments
are discussed in the beautiful review of Bauer, Spital, Yennie and
Pipkin\cite{BSYP78}. They summarize $f^2_\rho(q^2=m^2_\rho)/4\pi = 2.11\pm
0.06$ and  $f^2_\rho (q^2=0)/4\pi = 2.18\pm 0.22$,  as obtained from
experiments at the CEA, DESY, SLAC and Cornell. Real photon data at
energies from 3 to 10 GeV are used in the analysis.

No variation of $f_\rho(q^2)$ with $q^2$ is seen in the data! This seems to be
in strong disagreement with the consequences of the models of
Refs.~[42]-[48].  The survival of such models seems to depend
on finding a new way to account for the $\gamma p\to\rho^0 p$ data as well as
for data on many $\gamma$-nucleon and nuclear reactions.

Coon and Scadron\cite{CS94} use tadpole dominance to  argue that
$\rho^0 \omega$  exchange predicts an important charge symmetry breaking
nucleon-nucleon  potential.
In Feynman diagrams, tadpoles are  represented
by external lines which account for SU(2) and SU(3) symmetry breaking
effects
when inserted into a meson or baryonic line. The $\Delta I = 1$ tadpole
form contributes a term
$ H^3_{tad}=c^\prime H_3 = (m_u-m_d)(\bar u u - \bar d d)/2$
to the Hamiltonian. One can compute matrix elements  of the operator
$(\bar u u - \bar d d)$ using symmetry arguments. Then one can
account for the SU(2) and SU(3) violations of hadrons covering a wide
range of mass with only one mass-independent parameter.
Thus, in this picture,
the mixing matrix elements have no dependence  on the four-momentum
squared of the hadrons.

\begin{figure}
  \vspace{0.7in}
  \caption{Quark model of the $\rho^0$-$\gamma^*$ transition}
\end{figure}

Thus it is still reasonable\cite{M94,CS94}
to assume that $<\rho^0|H_{str}|\omega>$ has little
dependence on $q^2$.  Then charge symmetry breaking in the $^1$S$_0$ channel
is accounted for and there are many consequences, see Sects.~2.6, 2.7, 3.2 and
3.3.

A similar discussion regarding off-shell dependence  can be carried out with
respect to  $\pi^0$-$\eta$ mixing\cite{eta}. The presumed small value of the
$\eta$-nucleon coupling constant seems to make this effect much less
influential for nucleon-nucleon interactions than $\rho^0$-$\omega$
mixing. However, $\pi^0$-$\eta$ mixing could be  important in understanding
charge symmetry breaking in pion production reactions; see Sect.~3.3.  This
involves higher energy than elastic scattering,  so
the $\pi$ and the $\eta$ are on or near the
mass-shell and  a possible off-shell variation is less significant.
A new approach to $\pi-\eta$ mixing is discussed in Ref.~[54].

\begin{figure}
  \vspace{3.2in}
  \caption[test]{$q^2$ variation of ${1\over f_\rho}$.  The magnitude
 of $f_\rho$ has
been scaled to allow a comparison with the $q^2$ dependence of $<\rho^0|
H_{str}|\omega>$.}
\end{figure}

\vspace{0.5cm}
\noindent
{\em 2.4 Classification of Charge Dependent Nucleon-Nucleon Forces}

\vspace{0.2cm}
We have seen that some mechanisms contribute to $\Delta a_{CD}$ and others  to
$\Delta a_{CSB}$ and some to both. Moreover there are other observables
involving spin-dependent symmetry breaking effects. It is therefore useful to
characterize the charge dependence of nuclear  forces according to their
isospin dependence. The discussion of Henley and Miller\cite{HM79}
listed four classes.

Class (I): Forces which are isospin or charge independent. Such forces, $V_I$
obey $[V_I, \vec T]=0$ and thus have  the isoscalar form
\begin{eqnarray}
V_I= a +b\vec {\tau}(i)\cdot \vec {\tau}(j)
\end{eqnarray}
where a and b are reasonable isospin independent operators and i and j
label two nucleons.

Class (II): Forces which maintain charge symmetry but break charge
independence. These can be written in an isotensor form
\begin{eqnarray}
V_{II}=c [\tau_3(i)\tau_3(j)-{1\over 3} \vec{\tau}(i)\cdot \vec {\tau}(j)].
\end{eqnarray}
The Coulomb interaction leads to a Class II force as do the effects of
the pion mass difference in the one pion exchange interaction and
possible effects of charge dependent coupling constants.

Class (III): Forces which break both charge independence and charge
symmetry, but which are symmetric under the interchange $i\leftrightarrow
j$ in isospin space,
\begin{eqnarray}
V_{III}=d[\tau_3(i) +\tau_3(j)].
\end{eqnarray}
A class III force differentiates between $nn$ and $pp$ systems. However, it
does not cause isospin mixing in the two-body system, since
\begin{eqnarray}
[V_{III},T^2]\propto [T_3,T^2]=0.
\end{eqnarray}
This force vanishes in the $np$ system. The effects of the exchange of a
mixed  $\rho^0-\omega$ meson yield a significant class III force.
The effects of the quark mass difference that appear in the one-gluon
exchange interaction also contribute.

Class (IV): Class IV forces break charge symmetry and therefore charge
dependence; they cause isospin mixing. These forces are of the form
\begin{equation}
V_{IV} = e [\vec\sigma(i) - \vec\sigma(j)]\cdot\vec L[\tau_3(i)-\tau_3(j)]
+
f[\vec\sigma(i)\times\vec\sigma(j)]\cdot\vec L[\vec\tau(i)\times\vec\tau(j)]_3,
\end{equation}
where $e$ and $f$ are reasonable spin-independent operators. The $e$ term
receives contributions from $\gamma$, and $\rho \omega$ exchanges; while  $f$
is caused by the influence of the neutron-proton mass difference on  $\pi$ and
$\rho$ exchange. Charge dependence of the coupling constants  gives no Class IV
force.

Class IV forces vanish in the $nn$ and $pp$ systems, but cause spin-dependent
isospin mixing effects in the $np$ system.  As a result the analyzing power
of polarized neutrons scattered from unpolarized protons, $A_n(\theta_n)$,
differs from the analysing power of polarized protons scattered from
unpolarized neutrons, $A_p(\theta_p)$\cite{CHM78,AG78}. Measurements at
TRIUMF\cite{A86} and the IUCF\cite{K90}  compared scattering of polarized
neutrons to scattering of unpolarized neutrons from
an unpolarized, respectively
a polarized proton target.  Time reversal invariance relates the latter
measurement to $A_p$.  These analyzing powers pass through zero at one angle
$\theta_0$ for the energies of the TRIUMF  and IUCF experiments.
If $\theta_0$ for
polarized neutrons differs from  $\theta_0$ obtained for polarized protons,
then $\Delta\theta \equiv \theta_0(n) -\theta_0(p)\not= 0$ and charge symmetry
has been violated.  The results  of the two beautiful experiments  are
presented in terms of $\Delta A (= {dA\over d\theta}\Delta\theta)$, and
 are shown in
Fig.~9.  The calculations\cite{HTH87}  use the Bonn meson-exchange potential so
that all of the parameters governing the strong interaction are pre-determined.
(Other calculations are discussed in Ref.~[6].) The agreement between
theory and experiment is very good.  A pion exchange effect arising from the
presence of the $n$-$p$ mass difference in the evaluation of the vertex
function dominates the 477 MeV measurement\cite{MTW86}. The $\rho^0$-$\omega$
mixing term  has a significant but non-dominating influence at 183~MeV.

A new TRIUMF experiment, performed at 350 MeV, is discussed in Sect.~3.2.

\begin{figure}
  \vspace{4.0in}
  \caption[test1]{Measured values of $\Delta A\equiv A_n - A_p$ for $np$
elastic
scattering at 183 MeV (IUCF) and 477 MeV (TRIUMF).  The horizontal lines
represent theoretical predictions of  Ref.~\cite{HTH87}.}
\end{figure}

\vspace{0.5cm}
\noindent
{\em 2.5 The $^3$He-$^3$H Binding Energy Difference}

\vspace{0.2cm}

The ground state binding energy difference  $B(^3{\rm H}) - B(^3{\rm He}) =
764$ keV is a measure of the breaking of  charge symmetry\cite{O64}.
The  proton rich $^3$He nucleus
is less deeply bound because of the repulsive influence of  the Coulomb
interaction and other electromagnetic effects. Such effects  must be removed to
determine the strong interaction charge symmetry breaking.  The three body
system is the best for such evaluations because the most important
 electromagnetic terms can
be evaluated in a model independent way using measured electromagnetic form
factors\cite{F70}. Coon and Barrett\cite{CB87} used recent data to obtain
\begin{equation}
\Delta B(em) = 693\pm 19\pm 5~{\rm keV} ,
\end{equation}
where the first uncertainty is due to the determination of the form factors,
and the second to the small model dependence of some relativistic effects.
Similar values of $\Delta B(em)$ were obtained in Ref.~[63]. The
difference between 764 and 693 is about 71 keV, to be accounted for by charge
symmetry breaking of the strong interaction.  The use of a $\rho^0\omega$
exchange potential which reproduces $\Delta a_{CSB}$ yields about 90 $\pm$ 14
keV in good agreement with the experimental difference.  The errors allow some
room for other small effects such as $\pi\eta$ or $\pi\gamma$ exchanges. A
discussion of the precise value of the Coulomb  energy difference is given in
Ref.~[64].

\vspace{0.5cm}

\noindent
{\em 2.6 Nolen--Schiffer Anomaly and Other Nuclear Structure Effects}

\vspace{0.2cm}

The pattern of charge symmetry breaking seen for A=3 also occurs for the
mirror nuclei ($N,Z)$ = $(Z,Z+1)$ and $(Z+1,Z)$.
Charge symmetry predicts an equality between the binding energies, but
differences are observed.
It was first thought that electromagnetic, mainly Coulomb, effects
could account for all of the observed binding energy differences.
Evaluating the consequences of the electromagnetic effects is hindered by
the need to account for various correlation effects in the nuclear
wavefunction.
  Nolen and Schiffer\cite{NS69} made an
extensive analysis, finding that the
electromagnetic effects were not sufficient.
There was a residual effect due to charge symmetry breaking of the strong
interaction:
the neutron rich nuclei were  seen to be more deeply
bound (by about 7\%) than the proton rich nuclei.
Including additional
detailed nuclear structure effects reduced the number, but only
to a rather substantial 5\%. See the
reviews [6],[65] and Ref.~[66]. Negele also
suggested\cite{N71} that the charge symmetry violation in the two-nucleon
force could be responsible for the missing 5\% binding energy difference.

Since the early seventies
many authors have tried to better understand the anomalous
5\%, see the review~[6]. This was not easy because the
charge symmetry breaking of the
nucleon-nucleon force was very uncertain. An early
$\pi^-d\to nn \gamma$ experiment\cite{RMS75} obtained a scattering length
with an error of 1.3 fm.  This was large enough to say that the nn force
was either significantly more attractive or significantly more repulsive
than the pp  force. This was remedied by the PSI
measurements\cite{GTG87,OSc87} of the
neutron-neutron scattering lengths in the $\pi^-d\to nn \gamma$ reaction
which
indicated significantly more attraction for the nn force. %%
Additionally, the TRIUMF and IUCF np scattering experiments
increased the awareness of the fundamental mechanism for charge symmetry
breaking in the nucleon-nucleon interaction.

Thus the climate
was right for attacking the Nolen-Schiffer anomaly with a meson
exchange theory of the charge symmetry breaking
nucleon-nucleon potential when
Blunden and Iqbal\cite{BI87} took up the challenge.
As shown in Table~2, a good
account was obtained.

\pagebreak

\noindent Table~2. Blunden and Iqbal\cite{BI87}
treatment of the Nolen--Schiffer anomaly.
The first column identifies the nucleus and the single particle state.
The next two present the value of the non-electromagnetic contribution to
the binding energy
difference, as evaluated in
Ref.~[71] using the density matrix expansion DME or Skyrme
II interaction SKII.  The next two columns represent computed
binding energy differences, showing the total contributions  and the
individual effect of the
$\rho^0$-$\omega$ term, rescaled to use the
value of $<\rho^0|H_{str}|\omega>=-5200$~MeV$^2$.

\vspace{0.5cm}
\begin{center}
\begin{tabular}{|cc|cc|cc|}
\hline\hline
\multicolumn{2}{|c|}{A \,\, orbit} & \multicolumn{2}{c|}{Required CSB (keV)} &
 \multicolumn{2}{c|}{Calc. CSB (keV)} \\
&& DME &SkII   & total & $\rho^0-\omega$ \\
\hline
15& p$^{-1}_{3/2}$ & 250 & 190 & 210 & 182 \\
&p$^{-1}_{1/2}$ & 380 & 290 & 283 & 227 \\
\hline
17 & d$_{5/2}$ & 300 & 190 & 144 & 131 \\
& 1s$_{1/2}$ & 320 & 210 & 254 & 218 \\
& d$_{3/2}$ & 370 & 270 & 246 & 192 \\
\hline
39 & 1s$^{-1}_{1/2}$ & 370 & 270 & 337 & 290 \\
& d$^{-1}_{3/2}$ & 540 & 430 & 352 & 281 \\
\hline
41 & f$_{7/2}$ & 440 & 350 & 193 & 175 \\
& 1p$_{3/2}$ &380 &340 & 295 & 258 \\
& 1p$_{1/2}$ &410 &330 & 336 & 282 \\
\hline\hline
\end{tabular}
\end{center}

\vspace{0.5cm}
The total charge symmetry breaking
contribution to the binding energy difference is
in good  agreement with the amount required, which depends slightly on the
nuclear wave function.
Thus
charge symmetry  breaking effects in the strong interaction do account for
the missing binding energy difference, with the bulk
accounted for by the influence of $\rho^0$-$\omega$ mixing. Similar
results have been obtained in Refs.~[72] and [73].  Furthermore,
Krein, Menezes and Nielsen\cite{KMN93} included $\rho^0$-$\omega$ mixing in the
framework of the relativistic $\sigma,\omega$ (Walecka) model of nuclear
matter. Those   authors find a charge symmetry breaking effect of about the
right sign and magnitude, and an explanation of the same kind as that of
Refs.~[70]-[73]. Krein and collaborators\cite{KK94}  have recently
extended this  work to the case of finite nuclei,  with the same conclusion.
The Blunden-Iqbal explanation was criticized  in Ref.~[76] for using a
larger value of the $\rho$--nucleon coupling constant than that predicted by
vector dominance. However, the value is consistent with that of the Bonn
potential and  therefore seems well-constrained.

So far we have discussed explanations of the Nolen-Schiffer anomaly in terms of
vector mesons. But there is an entire class of
explanations$^{77-83,76}$ which examine the
scalar  effects of the nuclear medium which modify  the neutron-proton mass
difference, M$_n$-M$_p$.  Reducing  M$_n$-M$_p$ corresponds to
increasing the neutron attraction, which is needed  to explain the
Nolen--Schiffer anomaly.

\pagebreak

Several investigations were stimulated by the  work of Krein and
Henley\cite{GH89} who included effects of the nuclear medium
in the Nambu--Jona--Lasinio model\cite{dsb}
for chiral symmetry breaking.  Including a non-zero
nuclear density leads to a  Pauli principle  suppression of the
loop diagrams which generate the constituent mass.
This decreases the value of the quark condensates leading to a
partial restoration of
chiral symmetry in nuclei.
The result, as computed in a constituent quark model, is that  the
neutron-proton mass difference M$_n$-M$_p$ decreases
as the  nuclear density increases.
This is the kind of effect needed to account for the Nolen-Schiffer anomaly.
However,
when the constituent quark model
of a nucleon is  replaced by a chiral soliton model\cite{FCNBG91,MW91} or by  a
chiral model
with nucleon and meson degrees of freedom\cite{KMN92}, M$_n$-M$_p$
increases or remains the same  as the density is increased. These three papers
contradict the basic Krein and Henley result.

%Examination of the Krein--Henley equations, shows
%that including the  effects of unequal densities seen by the valence nucleon
%would  decrease the size of their effect, which was too big originally.

However, the Krein and Henley mechanism can be evaluated using QCD sum
rules$^{78-82,76}$.
The aim of such
treatments\cite{HHP91} is to express
the  neutron-proton mass difference to the
vacuum parameters $m_d-m_u$ and $\gamma=<\bar d d>/<\bar u u>-1$, in a
manner which avoids using an explicit model for the nucleon.
The use of QCD
sum rules yields results  that do tend to explain the Nolen--Schiffer anomaly.
This agreement may be illusory due to an important nuclear effect,  first
utilized by Williams and Thomas\cite{WT85}. The nuclear Coulomb  repulsion
pushes a valence proton away from the nuclear center. Thus a valence proton
sees a lower density than a valence neutron, i.e. $\rho_p<\rho_n$. This is a
substantial effect; $\rho_n =0.0667 fm^{-3}, \rho_p = 0.0594 fm^{-3}$ for the
valence nucleon of  $^{41}$Ca-$^{41}$Sc\cite{WT85}. Taking
this effect into account,
Fiolhais et al.\cite{FCNBG91} find a very small attraction for the
neutron even though  the neutron-proton mass difference increases with density
(for $\rho_n= \rho_p=\rho$).
Such
effects should be included in future QCD sum rule calculations. Note also that
a general criticism of such calculations is the
sensitivity  to  previously undetermined condensates,
see e.g. Ref.~[88].

The work of Saito and Thomas\cite{ST94} studies the Nolen--Schiffer anomaly
using their quark-meson  coupling model. In this mean-field model of  nuclear
matter non-overlap\-ping
nucleon bags are bound by the self-consistent exchange
of $\sigma,\omega$ and $\rho$ mesons. The effects of
self-consistent exchange of
$\sigma$ mesons combined with the quark mass difference
leads to different
bag energies for up and down quarks. A
qualitative, but not precise, explanation of the  anomaly is achieved.
 This work includes
the nuclear structure effect of Ref.~[87].

Thus there are two main mechanisms to explain the Nolen--Schiffer anomaly. The
vector effects of $\rho^0$-$\omega$ mixing; and, the scalar effects of the
medium
modifications (Pauli principle) of the nucleon mass. The first mechanism is
closely related to nucleon-nucleon scattering, Sect.~2.3,
and is  very traditional. It is
easy to relate this effect to the A=3 system for  which the nuclear wave
function is mainly irrelevant.
The second mechanism is   newer, more speculative, harder to
evaluate but has many implications. It is related to  explanations of the EMC
effect. Most likely, both mechanisms are present. Our view coincides with that
of Blunden and Iqbal, some 70\% or more of
 the anomaly is explained by $\rho^0$-$\omega$
mixing. This leaves plenty of room for  substantial scalar effects of the
nuclear medium. In any case,  the Nolen Schiffer anomaly is no longer a puzzle.
The quark  mass difference, whether by vector or scalar effects, is
responsible.

There are  other puzzles concerning  the energy
differences between heavier nuclei ($^{48}$Ca, $^{90}$Zr, $^{208}$Pb) and their
isobaric analog states. See e.g. the review\cite{AHKS72}.
The energy differences between these
nuclei and their isobaric  analog states are mainly due to Coulomb effects.
However, there is a  remainder to be attributed to charge dependent forces.
Naively, one would think that this remainder
should increase with the neutron excess. This was found not be
to be the  case\cite{ABVG80}. For example, in $^{48}$Ca, the sign of the
remainder  was less than for $^{41}$Ca. This puzzle was resolved by Suzuki,
Sagawa and Van~Giai\cite{SSVG93} who showed that it is necessary to use both
charge symmetry  and charge independence breaking  forces consistent with the
$NN$ data to compute the  relevant binding energy differences. The A dependence
seems now to be well understood. A remaining problem is that the purely
theoretical potentials do not give quantitative agreement. The theoretical
potentials used  in Ref.~[91] reproduce only about 70\% of $\Delta
a_{CD}$, as the $\gamma \pi$ term was not included.

Ormand and Brown have also included charge dependent nuclear  forces as input
to their  shell model calculations$^{92-94}$. An isospin
non--conserving interaction was used to understand the spectroscopic amplitudes
for the decay of T=3/2 states in  A=4n+1 nuclei (21$\le $ A $\le 37$) by proton
and neutron emission to the  T=0 ground states\cite{OB86} also to include
correction to the Fermi matrix element in $\beta$  decays\cite{OB85}. The
charge dependent interactions used are phenomenological, but  are in general
agreement with the results expected from nucleon-nucleon  scattering data. The
proton-neutron interaction  is about 2\% more attractive  than the average
of the proton-proton and neutron-neutron interactions.
Furthermore, including a charge symmetry
breaking interaction helps to  improve the  agreement between theory and
experiment.  It would be of interest to use charge dependent strong
interactions  as predicted by meson exchange theory, especially for the
contributions of  the $\pi^\pm$-$\pi^0$ mass difference and $\rho^0$-$\omega$
exchange.

\vspace{0.5cm}
\noindent
{\em 2.7 Charge Symmetry Breaking in Hypernuclei}

\vspace{0.2cm}

Hypernuclear binding energy differences  allow tests of charge  symmetry of the
$\Lambda$N interaction. The main source of information is the mirror pair
$^4_\Lambda H, \; ^4_\Lambda He$; see the reviews\cite{G75,G86}. Not much data
is available for other nuclei\cite{MNS90}. The difference between the $\Lambda$
separation energies of the  ground state\cite{J73}
\begin{eqnarray}
B_\Lambda(^4_\Lambda He)\equiv B(^4_\Lambda He)-B(^3He)=2.39\pm0.03 MeV
\end{eqnarray}
and
\begin{eqnarray}
B_\Lambda(^4_\Lambda H)\equiv B(^4_\Lambda H)-B(^3H)=2.04\pm0.04 MeV
\end{eqnarray}
provided evidence for a charge symmetry breaking component of the $\Lambda$- N
interaction\cite{DV64}.  These values must be corrected for Coulomb and other
electromagnetic effects to determine the strong interaction charge symmetry
breaking. A complete four-body calculation is required to do this, a folding
model of the $\Lambda$ nucleus interaction is not sufficient\cite{GL81}.

Extensive work on the $\Lambda$N interaction, including the effects of charge
symmetry breaking, was done by the Nijmegen group\cite{N73} which takes the
most important source to be $\Lambda$-$\Sigma^0$ mixing (Sect.~2.2).  This
allows
a  $\pi^0$ to couple to the physical $\Lambda$, so the charge  symmetry
breaking potential has the range of one pion exchange and  consists of a
spin-spin and tensor interaction. The $\Sigma$ is only  80~MeV heavier than the
$\Lambda$, so  coupling to the $\Sigma$N channel is important and  the mass
differences between the  different $\Sigma$ charge states also contribute.
Comprehensive
fits to data  for $\Sigma$-$p$ reactions led this group to predict the
charge symmetry breaking  of the $\Lambda$N interaction, even though there are
no data for that channel. Gibson and Lehman\cite{GL79} made model
calculations using separable potentials fitted to the scattering  length and
effective ranges of the Nijmegen model D potential.  This
was sufficient to account for
the observed charge symmetry breaking of the ground states.  Gibson\cite{BG94}
emphasized the importance of the $\Sigma$ mass differences in computing the
level orderings of the $^4$He ground and excited states.  However, it is
reasonable to assume that the effects of $\Lambda$-$\Sigma^0$ mixing and
$\rho^0$-$\omega$ mixing are far more important for the $\Lambda N$ charge
symmetry breaking interaction.

Bodmer and Usmani\cite{BU85} considered also the charge symmetry  breaking of
the first excited states, using extensive variational calculations. The
$\Lambda$ separation energies of the lowest 1$^+$ excited states are\cite{B79}
\begin{eqnarray}
B_\Lambda(^4_\Lambda He^\ast)
\equiv B(^4_\Lambda He^\ast)-B(^3He)=1.24\pm0.06 MeV
\end{eqnarray}
and
\begin{eqnarray}
B_\Lambda(^4_\Lambda H^\ast)\equiv B(^4_\Lambda H^\ast)-B(^3H)=1.00\pm0.06~MeV.
\end{eqnarray}
They determined the Coulomb contributions to improved accuracy; the differences
of B$_\Lambda$  for the ground and excited states were found  to be
0.40$\pm$0.06 and 0.27$\pm$ 0.06  MeV. These results  could be described by a
charge symmetry breaking potential which effectively is independent of
spin, in contrast with the one-pion-exchange effect. Thus, the potentials
of Ref.~[98] do not describe the charge symmetry breaking when one
considers both the ground and excited states.  This conclusion relies upon
the ability of variational methods to handle the four body problem.  Bodmer
and  Usmani used a very detailed correlated wave function. They tested
their procedure by applying it to $^3$He,  obtaining  excellent agreement
between their variational wave function and the exact Green function
results. Thus the procedure seems adequate. It would  still be desirable to
confirm the Bodmer-Usmani results with an independent calculation.

Thus the need to understand why the $\Lambda$N charge  symmetry breaking
interaction is  approximately independent of spin  is an unsolved problem,
so we make a  suggestion. The effects of exchange of a mixed
$\rho^0$-$\omega$ meson have  been ignored since the very early work of
Downs\cite{D66} who argued that cancellations take place when  the effects
of mixing between the $\rho,\omega$
and  $\phi$ mesons are all included. But this is based on early values of
meson-baryon coupling constants, and on radial potentials with a  Yukawa
form instead of the exponential form obtained when two  meson  propagators are
involved. It is certainly time to take a  new look at the role of meson mixing
in the $\Lambda$N interaction. In particular, the exchange of a mixed
$\rho^0$-$\omega$ meson leads to a  substantial central interaction.

\vspace{0.5cm}
\noindent
{\bf 3. New experimental information}

\vspace{0.5cm}
Effects of charge independence and charge symmetry breaking have been  sought
with many different projectiles and targets. These experiments are
intrinsically challenging because of the small nature in most instances of the
effects being  sought. Nevertheless, substantial progress has been achieved.
%This is discussed below, concentrating on recent work.

\vspace{0.5cm}
\noindent
{\em 3.1 Low-Energy Nucleon-Nucleon Scattering}

\vspace{0.2cm}

The earliest evidence for the breaking of charge independence  in
nucleon-nucleon scattering was  the inequality of the N-N scattering lengths.
In particular, charge independence breaking occurs in  the difference ${1\over
2} (a^{N}_{nn} + a^{N}_{pp}) - a^N_{np} = (5.7 \pm 0.3)$ fm, where
$a^{N}_{nn}$, $a^{N}_{pp}$, and $a^{N}_{np}$ are given  in Table~3.  This
difference of 5.7 $\pm$ 0.3 fm can be explained as resulting predominantly from
the mass difference between the neutral and charged pions; see Sect.~2.3.  The
quantities $a^{N}_{pp}$, $a^{N}_{nn}$ and $a_{np}^N$ are  $pp$, $nn$ and $np$
scattering lengths, corrected for Coulomb effects and  vacuum polarization.

\vspace{0.2cm}
%%%PLEASE DO NOT REPLACE \CITE{ABC} BY A SPECIFIC NUMBER
\begin{center}
Table~3 \\
Low energy nucleon-nucleon scattering observables
\end{center}
\renewcommand{\arraystretch}{1}
\begin{tabular*}{16cm}{@{\extracolsep{\fill}}cccccc}
\hline\hline
& & & & & \\
      & $nn^{67,68,69}$    & $nn^{N}$\cite{MNS90} &
$np^N$\cite{ODu83} & $pp$\cite{NIM} & $pp^N$\cite{MNS90}  \\
\hline
& & & & & \\
$a$(fm) & -18.45$\pm$0.32 & -18.8$\pm$0.3 & -23.748$\pm$0.009&
-7.8063$\pm$0.0026 &    -17.3$\pm$0.4   \\
$r$(fm) & 2.80$\pm$0.11 & 2.75$\pm$0.11 & 2.75$\pm$0.05 & 2.794$\pm$0.0014 &
2.85$\pm$0.04 \\
\rule[-.1cm]{0cm}{0.5mm} \\
\hline\hline
\end{tabular*}

\vspace{0.5cm}

Charge symmetry breaking is a smaller effect than charge independence and
therefore is more subtle.   Its influence can be seen through a difference of
the $pp$ and $nn$ scattering  lengths, but  obtaining a precise value has
been
difficult. The ability to  remove Coulomb effects  from the experimentally
observed pp scattering length depends on knowing  the  short range part of the
nucleon-nucleon interaction\cite{SAUER}. Nowadays,
quark models can be used to limit the
uncertainties in the short-distance strong interaction; see  the
review\cite{MNS90} and Ref.~[106]. Progress is possible.

It is the lack of a free neutron target  that causes the most
difficulties in observing neutron-neutron  scattering and its difference
from proton-proton scattering. Proposals to
scatter neutrons from neutrons have appeared from time to time. But no real
experiment has ever been reported in the scientific literature.
The most reliable determinations of $a_{nn}$
as well as the effective range $r_{nn}$ occur in three-body reaction studies
with only two strongly interacting particles (the two neutrons) in the final
state. Consequently  $a_{nn}$ and $r_{nn}$ are mainly deduced in studies of the
reaction $\pi^{-}d\rightarrow\gamma nn$.   The result is a small inequality of
the nn and pp scattering  lengths, with the nn scattering length
($a^N_{nn}=-18.8\pm 0 .3$ fm\cite{MNS90}) slightly more negative than the pp
scattering length ($a^N_{pp}=-17.3\pm0.4$ fm\cite{MNS90}), giving a difference
of $a^N_{pp}-a^N_{nn}=-1.5\pm 0.5$ fm. A 1 fm difference  between the
scattering lengths corresponds to only about a part in 200  difference  between
the  potentials. It is necessary to find other  manifestations of charge
symmetry breaking to establish its  existence.

We  next discuss the use of
the deuteron  breakup reactions nd$\to$nnp and pd$\to$ppn
in determining nucleon-nucleon scattering parameters.
The  three nucleon final states must be analyzed with
Faddeev calculations in order to extract the low-energy nucleon-nucleon
scattering parameters.
The treatment of Coulomb effects in Faddeev calculations
is extremely complicated and cannot be done without approximations;
see eg. the recent review\cite{Gloe}.
For a recent report on including the Coulomb interaction in $Nd$ breakup
calculations see ref.~[107].
A precise direct comparison between nn observables (from nd) and pp
observables (from pd) is not yet possible.

It is more reasonable to attempt to
determine the nn scattering parameters from the nd breakup reaction.
However, even though a great deal of effort has gone
into understanding the three nucleon final state reactions in a quantitative
way\cite{Gloe,SAT89},
  we continue to doubt their use in obtaining  precise values of
of $a_{nn}$ and
$r_{nn}$.
A recent example of such a determination is the
measurement of $^{2}H(n,nnp$) at 10.3 and 13.0~MeV\cite{KG93},
which contains several advances.
Comparisons
between experiment and modern
theory in various regions of phase space show
good agreement.
Furthermore, these authors test their method by trying to extract
the known value of a$_{np}$.
Including a charge dependent improvement to the Paris potential
clearly  improves the agreement between theory and experiment. Thus they
clearly demonstrate that their
method can certainly observe a 5-7 fm effect in the scattering
length. However, the level of precision required to learn about charge
symmetry breaking is about 0.5 fm or better.
We doubt this precision is possible now.
Even in the kinematic region where the np final state interaction
dominates, the agreement is not excellent; and there are significant
disagreements between theory and experiment in other regions
of phase space.
There are other obstacles.
It has been reported that the extraction of $a_{nn}$ from kinematically
complete nd breakup data in state-of-the-art analyses is sensitive to the
choice of the nucleon-nucleon potential\cite{TORN}.  Even if one starts
with  realistic
nucleon-nucleon interactions, it is often convenient to
make approximations, such as using a single  separable potential,
to  incorporate
these interactions in  the Faddeev formalism.
Furthermore, the extracted value of a$_{nn}$ is sensitive to whether or
not the effects of three body forces are included in
the analysis\cite{Gloe}.
For all these reasons, we
question the present ability to determine the
precise value of a$_{nn}$ from
the nn final state interaction  region in the nd break-up reaction.

A new experimental effort to measure the low-energy nn scattering parameters
is underway at TRIUMF\cite{MAK661}.  The experiment is a three-fold coincidence
measurement of the $\pi^{-}d \rightarrow \gamma nn$ reaction with stopped
pions.  It is anticipated that the nn scattering length $a_{nn}$ and
effective range parameter $r_{nn}$ can be determined to an accuracy comparable
to or better than obtained previously in the PSI
measurements\cite{GTG87,OSc87}.

\vspace{0.5cm}
\noindent
{\em 3.2 $np$ Scattering}

\vspace{0.2cm}

Charge symmetry leads to the complete separation of the isoscalar and isovector
components of the $np$ interaction.  This in turn leads to the equality of the
differential cross sections for polarized neutrons scattering from unpolarized
protons and vice versa.  As a result $A_n(\theta) \equiv A_p(\theta)$ where $A$
denotes the analyzing power and where the subscript represents the polarized
nucleon.  A nonvanishing asymmetry difference is directly proportional to the
isotopic spin singlet-triplet, spin singlet-triplet mixing amplitude and
therefore direct evidence of a charge-asymmetric interaction,
anti-symmetric under the interchange of nucleons 1 and 2 in isotopic spin
space; see Sect.~2.4.

The scattering matrix for $np$ elastic scattering can be expressed in terms of
the  formalism of LaFrance and Winternitz\cite{LF80} as
\begin{eqnarray}
M(\vec{k}_f,\vec{k}_i) = {1\over
2}\lbrace(a+b)+(a-b)(\vec{\sigma}_1\cdot\hat{n})(\vec{\sigma}_2\cdot\hat{n}) +
(c+d)(\vec{\sigma}_1\cdot\hat{m})(\vec{\sigma}_2\cdot\hat{m}) \nonumber \\
+(c-d)(\vec{\sigma}_1\cdot\hat{\ell})(\vec{\sigma}_2\cdot\hat{\ell}) +
\epsilon(\vec{\sigma}_1 + \vec{\sigma}_2)\cdot\hat{n} + f(\vec{\sigma}_1 -
\vec{\sigma}_2)\cdot\vec{n}\rbrace.
\end{eqnarray}
Here $\hat{\ell}, \, \hat{m}$ and $\hat{n}$ are unit vectors given as
\begin{eqnarray}
\hat{\ell} = {\vec{k}_i + \vec{k}_f\over |\vec{k}_i + \vec{k}_f |}; \,\,
\hat{m} =
{\vec{k}_f - \vec{k}_i\over |\vec{k}_i + \vec{k}_i |}; \,\, \hat{n} =
{\vec{k}_i
\times \vec{k}_f\over |\vec{k}_i \times \vec{k}_f |};
\end{eqnarray}
with $\vec{k}_i$ and $\vec{k}_f$ the initial and final state center-of-mass
nucleon momenta.  The amplitudes $a, \, b, \, c, \, d, \, e$ and $f$ are
functions of center-of-mass energy $E$ and scattering angle $\theta$.  Written
explicitly, the difference in the analyzing powers
\begin{eqnarray}
\Delta A(\theta) \equiv A_n(\theta) - A_p(\theta) = {2\over \sigma_0}
{\cal R}e(b^*f),
\end{eqnarray}
is proportional to $f$. The quantity $\sigma_0$ is the differential cross
section for the scattering of unpolarized neutrons from unpolarized protons.

Experimental considerations show that the next least difficult  quantity to
measure is the difference in the spin-correlation parameters $C_{xz}(\theta)$
and $C_{zx}(\theta)$.  The correlation parameter $C_{xz}$  requires  the
projectile spin to be polarized  transverse to the beam direction in the
scattering plane and the target spin polarized along the incident beam
direction, while for $C_{zx}$ the reverse holds.  Charge symmetry leads to the
equality of $C_{xz}(\theta)$ and $C_{zx}(\theta)$, but if charge symmetry is
broken then one would be able to measure a difference
\begin{eqnarray}
\Delta C(\theta) \equiv C_{xz}(\theta) - C_{zx}(\theta) = {2\over \sigma_0}
Im(c^{*}f).
\end{eqnarray}
Due to the intrinsic difficulties of excluding extraneous polarization
components in both the incident beam and the scattering target,
measurements of $\Delta C \equiv
C_{xz} - C_{zx}$ have not yet been attempted.  Other observables are defined
by LaFrance
and Winternitz\cite{LF80}.

The first measurement of charge symmetry breaking  in $np$ elastic scattering
was performed at TRIUMF\cite{A86}.  The measurement of $\Delta A \equiv A_{n} -
A_{p}$, at the zero-crossing angle of the average  analyzing power, at an
incident neutron energy of 477 MeV, yielded $\Delta A$ = (47 $\pm$ 22 $\pm$
8)$\times$10$^{-4}$,  a little over two standard deviations effect.  More
recently the results of a similar experiment at a neutron energy of 183 MeV
performed at IUCF have been reported\cite{K90}.   The measured value of $\Delta
A \equiv A_{n} - A_{p}$, averaged over the angular range 82.2$^{\circ}$ $\leq
\theta_{cm} \leq$ 116.1$^\circ$ over which $<$ $A(\theta)$ $>$ averages to
zero, is (33.1 $\pm$ 5.9 $\pm$ 4.3) $\times$ 10$^{-4}$, where again the first
error represents mainly the statistical uncertainty and the second error the
systematic uncertainty.  The latter result differs from zero by 4.5 standard
deviations (see Fig.~9).  It differs from the value expected from the
electromagnetic spin-orbit interaction by 3.4 standard deviations.  This
difference represents the strongest experimental evidence to date of
charge symmetry breaking in the nuclear interaction.

There are difficulties in extracting an angular distribution of $\Delta
A(\theta)$.  This follows directly from the expression for the difference in
the asymmetries for beam and target polarized, respectively, or
\begin{eqnarray}
\epsilon_b(\theta) - \epsilon_t(\theta) = \Delta A(\theta)(P_b + P_t)/2
+ < A(\theta) > (P_b - P_t),
\end{eqnarray}
pointing to the need for calibration of the beam and target polarizations
($P_b$ and $P_t$) with an accuracy unattainable at present.  In the analysis of
the IUCF experiment this difficulty was overcome by adjusting the ratio of
($P_b/P_t$) until the error-weighted rms value of $\Delta A(\theta)$ over the
angular range of the experiment reached minimal variance.  Following this
procedure a twelve point angular distribution was obtained, see Fig. 10.
The procedure
does not work at 477 MeV where $\Delta A (\theta)$ and $< A (\theta) >$ have
zero-crossing angles in close proximity and consequently the angular
dependencies are no longer orthogonal.  If the theoretical calculations were
precise in their predictions of the zero-crossing angle of $\Delta
A(\theta)$ one could in principle also determine $\Delta P = P_b - P_t$ and
consequently the angular distribution of $\Delta A(\theta)$ would follow.

In general, the measured analyzing power differences of the IUCF and TRIUMF
experiments are well reproduced by theoretical predictions based on meson
exchange potential models, which indirectly incorporate quark level effects.
The calculations include contributions from one photon exchange (the magnetic
moment of the neutron interacting with the current of the proton), from the
neutron-proton mass difference affecting charged one $\pi$,
$\rho$ and $\omega$ exchange,
and from the more interesting isospin mixing $\rho^0$-$\omega$
meson exchange.  Some other smaller effects (like two $\pi-$exchanges not
included in $\rho-$exchange) have
also been evaluated\cite{JAN92}.  The effects of
$\pi \gamma$ exchanges have not yet been calculated for the two experiments
under discussion.  The theoretical results indicated by the solid and dashed
lines in Fig.~9 are based on a momentum space version of the Bonn
nucleon-nucleon potential\cite{HTH87}.  Note that the first two contributions
mentioned suffice to give a theoretical prediction in agreement with the TRIUMF
result.  This is because at 477~MeV the effect arising from $\rho^0$-$\omega$
mixing crosses zero close to the zero-crossing angle of $<A(\theta)>$.
Reproducing the IUCF result at 183~MeV with present calculations  requires
inclusion of the $\rho^0$-$\omega$ meson mixing contribution, an
approximately two standard deviation effect.

\vspace{4.8in}
\baselineskip=12pt
\noindent Figure 10: Comparison of the measured $\Delta A(\theta)$ angular
distribution at 183 MeV [IUCF] with calculations based on Ref.~[59] and
Ref.~[114], which display different distorting potentials and different
$\rho NN$ and $\omega NN$ coupling constants.  Experimental data and
theoretical curves have been subjected to a $\Delta A(\theta)$ variance
minimization procedure.  Total $\chi^2$ values for each curve include
statistical errors only (see [58]).

\pagebreak
\baselineskip=14pt
Figure 10 shows
the calculations of Holzenkamp, Holinde and
Thomas\linebreak[4]
(HHT)\cite{HTH87} and of Beyer and Williams (BW)\cite{BW88}. The
differences  reflect different
$\rho$NN and $\omega$NN coupling constants.  HHT use larger coupling constants,
as determined by the the requirement that their one boson exchange potential
model reproduce the
charge symmetric data.
Thus the better agreement
with the data provided by calculations employing the stronger coupling
constants is due to internal consistency among the ingredients of the HHT
calculations.  The charge symmetry breaking effects are evaluated using the
same coupling constants that determine the charge-symmetric interaction.
A calculation, with a similar consistency, has  recently been made by Iqbal and
Niskanen\cite{IN88}.  However, there is controversy regarding the role of
$\rho^0$-$\omega$  mixing, recall Sect.~2.3.  The
theoretical predictions of Iqbal and
Niskanen include contributions from one
photon exchange, the np mass difference affecting
charged $\pi$ and $\rho$ exchange, $\rho^0$-$\omega$ mixing,
and two $\pi$ exchanges
not included in $\rho$ exchange.  The
np scattering wave function was derived using the
np phases from a charge independent phase shift analysis\cite{RAALDR80}.  The
effects of
inelasticity amount to about 10\% at 800 MeV but are vanishing small
below\cite{JANSEV92}.

A new measurement at
350~MeV has been made at TRIUMF (E369)
(Ref.~[118]), with data taking
completed in the spring of 1993,
to delineate the various contributions to \csb.
This experiment
is very similar to the
earlier TRIUMF measurement at 477  MeV\cite{A86}.  The 350~MeV neutron beam was
produced using the ($p,n$) reaction on deuterium.  The proton beam had an
intensity of about 2$\mu A$ and a polarization of about 0.70 and was incident
on a 0.21 m long LD$_2$ target.  The energy, the polarization, the position and
direction of the proton beam were monitored throughout the experiment and
controlled (in the case of position and direction) using a feedback system
coupling two sets of split-plate secondary electron emission monitors (which
determined the median of the intensity distribution) with steering magnets
upstream in the beam transport line.  At the two sets of split-plate secondary
emission monitors the beam position was kept fixed with a standard deviation of
$\leq$ 0.05~mm in both $x$ and $y$ intensity profiles.  The beam energy
monitor,
based on range determinations, allowed the beam energy to be kept constant with
a standard  deviation of less than 30~keV (through minute changes in rf of the
cyclotron and stripper foil position).  The polarization was transferred from
the proton to the neutron by making use of the large sideways to sideways
polarization transfer coefficient $r_t$ (-0.88 at 364~MeV).  Required rotations
of the polarization directions were obtained by a solenoid magnet (for the
proton polarization direction) and a combination of two dipole magnets (for the
neutron polarization direction).  The 9$^\circ$ neutron beam passed a 3.3~m
long, tapered steel collimator before impinging on a frozen spin type polarized
proton target (containing butanol beads) positioned at 12.85~m from the center
of the LD$_2$ target. Typical polarizations were 0.80 or higher.  Scattered
neutrons were detected in the angle range 24.0$^\circ$ to 42.4$^\circ$ in large
area scintillation counters, while the recoil protons were observed in
scintillation counter/wire chamber telescopes nominally centered at 53$^\circ$.
The detection apparatus had reflection symmetry about the neutron beam axis to
increase the event rate and to allow with reversals of the beam and target
polarization directions certain systematic errors to be cancelled (a
three dimensional picture of the detector setup is shown in Fig.~11).  At
the zero-crossing angle all systematic errors, except those due to background
corrections, are eliminated to second order in an expansion, in the error
contributions. Further experimental details can be found in [119].

\vspace{4.5in}
\noindent Figure 11: Three dimensional view of the detection apparatus of
the TRIUMF 350 MeV experiment measuring CSB in $np$ elastic scattering.

\vspace{0.5cm}
To select elastically scattered $np$ events, proton and neutron tracks were
reconstructed and their energies were calculated from time--of--flight.
Four kinematic variables, opening angle, coplanarity angle, kinetic energy sum
and horizontal momentum balance, were formed and momentum dependent chi--square
cuts were applied.  Scattering asymmetries were calculated for the selected
events, which included a small contribution of quasi--elastic ($n,np$)
background events.

Experimental parameters were studied to understand the source of systematic
errors.  These parameters include  beam energy, proton and neutron
polarizations, split--plate secondary electron emission monitor asymmetries,
neutron profile parameters and frozen spin target parameters.  Corrections were
made for the contribution of quasi--elastic background and for the average
neutron beam energy difference due to the energy and polarization correlation
of the neutron beam production reaction $D(\vec{p},n)2p$.  The
frozen spin target with the the butanol beads replaced by
carbon beads was used to
study the background contribution.  It was determined that 2.5\% of the
selected events,
with $\chi^2 \, \leq$ 6 cuts in the four kinematical variables,
were from background.
The opening angle distributions of the carbon target
data were normalized to those of
the butanol target data by matching the tails of the distributions.
Neutron beam energy and the energy dependence of the
effective spin transfer coefficient were studied with a Monte Carlo simulation.
The difference of effective average neutron beam energy of polarized and
unpolarized beam was calculated and a correction was obtained.

A preliminary result for the difference of the zero--crossing angles is
$\Delta\theta_{cm} = 0.48^\circ \pm 0.08^\circ(stat.) \pm 0.08^\circ(syst.)$
based on fits of the asymmetry curves over the angle range $55.8^\circ \leq
\theta_{cm} \leq 85.4^\circ$.  With $dA/d\theta_{cm} = (-1.35 \pm 0.05)\times
10^{-2}
{\rm deg}^{-1}$, as determined from the measured asymmetries with the polarized
target, the value of $\Delta A \equiv A_n - A_p$, where subscripts denote
polarized nucleons, is $[65 \pm 11(stat.) \pm 11(syst.)] \times 10^{-4}$.
More extensive data analysis is in progress.  It is expected that with all data
analyzed a statistical accuracy of about $\pm 0.04^\circ$ in zero--crossing
angle difference or $6 \times 10^{-4}$ in $\Delta A$ as well as a more definite
systematic error will be obtained\cite{JZSJS93}.

\vspace{0.5cm}
\noindent
{\em 3.3 Pion Production Experiments}

\vspace{0.2cm}
\noindent
a) $np \rightarrow d \pi^0$

\vspace{0.1cm}
Another experiment that tests the effects of class IV interactions is the
measurement of the forward-backward asymmetry in the reaction  $np\rightarrow
d\pi^0$.  In the absence of such interactions  isospin is conserved and
consequently the angular distribution is symmetric about 90$^\circ$ in the
center-of-mass.  A nonvanishing forward-backward asymmetry difference in the
deuteron angular distribution

\begin{eqnarray}
A_{fb} \equiv {\sigma(\theta) - \sigma (\pi - \theta)\over \sigma (\theta) +
\sigma(\pi - \theta)}
\end{eqnarray}

\noindent
is direct evidence of a charge-asymmetry in the pion production interaction,
which is anti--symmetric under the interchange of nucleons 1 and 2 in
isotopic spin space.

The first calculations of $A_{fb}$ by Cheung, Henley and Miller\cite{CHM78}
show an energy dependence of $A_{fb}$ that suggests a large negative value
below 300 MeV.  Recent calculations\cite{NS88} give an angle integrated value
of approximately -50 $\times$ 10$^{-4}$ for $A_{fb}$ near 280 MeV with $\pi^0-
\eta$ and $\pi^0- \eta '$ mixing contributing almost an order of magnitude more
than one photon exchange, the neutron-proton  mass difference affecting charged
one $\pi$- and $\rho$-exchange, and  $\rho^0$-$\omega$ mixing.  These
contributions are strongly dependent on the meson mixing matrix elements and on
the $\eta NN$ coupling constant.  The   recent studies of van Kolck\cite{VK93}
display the importance of
\csb \, in $\pi^0$-nucleon scattering. This is because
the charge symmetric scattering length $a_0$ is proportional to $m_d+m_u$,
which
 would vanish in the limit of
complete chiral symmetry. This allows the usually small
\csb \, correction, proportional to $m_d-m_u$, to be comparatively large.
The $\pi^0$-nucleon scattering influences pion production through its
contribution to the rescattering matrix element, but this effect has not
yet been evaluated.

While $A_{fb}$ has been measured for a range
of energies, the large uncertainties of these measurements make them rather
inconclusive$^{122-125}$.
A new measurement of $A_{fb}$ at 281 MeV is in a preparatory stage at
TRIUMF\cite{KO704}.   Using a broad range magnetic spectrometer set at zero
degrees, the full angular distribution of the recoil deuterons will be measured
in a single momentum setting, eliminating many systematic uncertainties.
Remaining systematic uncertainties will be suppressed by measuring the response
of deuterons with the same momentum and trajectory from  $pp\rightarrow
d\pi^+$,  which must be symmetric about 90$^\circ$ in the centre-of-mass.  The
group's intention is  to measure $A_{fb}$ with a precision of approximately
$\pm$7 $\times$ 10$^{-4}$.

Charge symmetry also imposes constraints on  the vector and tensor
polarizations of the deuteron if the reaction proceeds with unpolarized
neutrons incident on an unpolarized proton target.  It is to be noted that the
vector polarization $P_y$ and the tensor polarization component $P_{xz}$ are
odd functions in $\theta$, while the tensor polarization components $P_{xx}, \,
P_{yy},$ and $P_{zz}$ are even functions in $\theta$.  Obtaining quantitative
information about deviations from these symmetry relations will certainly be
much more  difficult than measuring the forward--backward asymmetry in the
differential cross section angular distribution.

\vspace{0.5cm}
\noindent
b) $dd \rightarrow ^4$ He $\pi^0$

\vspace{0.1cm}
In self--conjugate systems (which are characterized by $T_3 = 0$) a
charge--symmetric Hamiltonian cannot connect states which differ in isospin by
one unit.  Consequently the reaction $dd \rightarrow$ $^4$He$\pi^0$ is
forbidden if charge symmetry holds, so observing
a non-zero differential cross
section proves that charge symmetry  is broken.
The forward angle center-of-mass differential cross section at an
incident energy of 600~MeV was estimated to be 0.2  pb/sr\cite{CC84}
mainly from a mechanism in which $\pi^0$-$\eta$ and $\pi^0$-$\eta'$
mixing is enhanced by the
resonant (3,3) pion-nucleon interaction.
Coon and Preedom\cite{CP86} deduced a
center of mass differential cross section for $dd \rightarrow ^{4}$He $\pi^0$
at 1.95 GeV of 0.12 $\pm$ 0.05 pb/sr. Their mechanism is the
dd$\to ^4$He$\eta$ reaction followed by external $\eta-\pi^0$ mixing.
Their value for the $dd \rightarrow$$^4$He$\pi^0$ cross section was obtained
using a measured value for the reaction $dd
\rightarrow$ $^{4}$He$\eta$ at 1.95~GeV and at $\theta_{cm}$ = 146$^\circ$ of
$d
\sigma/d \Omega$  = 0.25 $\pm$ 0.10 nb/sr\cite{JB85}.

An extensive search for the reaction $dd \rightarrow$ $^{4}$He$\pi^0$ has been
made at Saturne.  The most recent  experiment\cite{JB87} reported a limit of
0.8 pb/sr (68\% C.L.) at an incident deuteron energy of 800 MeV and around
100$^\circ$ cm.  In a later  publication\cite{LG91} about the same experiment a
differential cross section was reported of $d \sigma / d \Omega$ =
0.97$\pm$0.20$\pm$0.15 pb/sr at $\theta_{cm}$ = 107$^\circ$ at an incident
energy of 1.10 GeV, with the first error representing the statistical
uncertainty and the second error the systematic uncertainty.  This result would
be the first to establish a differential cross section for $\pi^0$ production
in the reaction $dd \rightarrow$$^4$He$\pi^0$.  The reaction
$dd\rightarrow$$^4$He$\gamma$ was measured simultaneously with $\pi^0$
production, leading to a differential cross section of $d \sigma/d \Omega$ =
0.82$\pm$0.18$\pm$0.10 pb/sr at $\theta_{cm}$ = 110$^\circ$.
We stress that
the experiment is extremely difficult to execute and to analyze.
In particular,
establishing unambiguously that the two photon events originated indeed in a
$\pi^0$ from $dd \rightarrow$ $^{4}$He$\pi^0$ is a major problem.

Wilkin\cite{W94} has used the measured $dd\rightarrow$$^4$He$\pi^0$
cross section
along with an extrapolation of the $dd\rightarrow$$^4$He$\eta$ data
and the Coon and Preedom external mixing model to determine an
$\eta$-$\pi^0$ mixing angle which is compatible with other determinations
reviewed in Ref.~[6]. A consequence is significant ($\approx 10$
\%)  violations of
charge independence in the ratio of the $pd\rightarrow$$^3$He$\pi^+$ and
$nd\rightarrow$$^3$H$\pi^0$ cross sections at energies near
the $\eta$ production
threshold.

\vspace{0.5cm}
\noindent
{\em 3.4 Pion Scattering Experiments}

\vspace{0.2cm}
Many attempts have been made to compare $\pi^-$ and $\pi^+$ scattering from
self-conjugate nuclei (isospin singlets).  The first such an attempt was a
comparison between the $\pi^-$ and $\pi^+$ total cross sections for
deuterium\cite{EP78}. The authors
used high statistics data with two targets of different lengths.
The necessary Coulomb correction  was simplified by the dominance of the
single scattering term. The results
were analyzed
in terms of the energies and widths of the various $\Delta$ resonances.
Some very reasonable results were obtained.  The extracted
$\Delta$ mass differences
are consistent with those expected from the quark model.

The overriding uncertainty in many comparisons is in the removal of all effects
related to the Coulomb interaction.   Even if this is done precisely,
many experiments are not sufficiently
precise, in  both statistics and systematics, to warrant a
quantitative deduction of charge symmetry breaking
effects. The early work on comparing the
$\pi^\pm d$ total cross sections was followed by a series of
measurements of the differences between the  $\pi^-d$ and  $\pi^+d$
differential cross sections.  Here an asymmetry parameter can be defined as,
\begin{eqnarray}
A_\pi = {(d \sigma/d\Omega)_{\pi^-} - (d\sigma/d \Omega)_{\pi^+}\over (d\sigma
/
d\Omega)_{\pi^-} + (d\sigma / d\Omega)_{\pi^+}}.
\end{eqnarray}

Multiple-scattering causes important effects in the angular distribution,
except in the forward direction (where  the Coulomb interaction is most
important). This means that
accounting for Coulomb interaction effects is a difficult
task in a three-body Faddeev or any other calculation. Once this is done,
 the differences may be related to the
differences in energies and widths of the various $\Delta$ resonances, as
presented in Ref.~[134].   However, as stated for instance by K\"{o}hler
et al.\cite{MK91}, the errors on the  values of $A_\pi$ are in most cases of
about the same size as the values of $A_\pi$ themselves.  Therefore, it is
difficult to extract evidence for \csb.

A phase shift analysis of $\pi^-$-$^{4}$He and $\pi^+$-$^{4}$He elastic
scattering data was made by Brinkm\"{o}ller and  Schlaile\cite{BS93} following
the measurement of differential cross section angular distributions at a series
of energies between 90 and 240~MeV\cite{BB91}.  For energies where both $\pi^-$
and $\pi^+$ measurements exist, one can compare the phase shift solutions for
$A_\pi$ to the experimental charge asymmetry $A_\pi$.  One finds that the
treatment of Coulomb effects, as done in the phase shift analysis as the only
origin for \csb, is adequate.  Once more, intrinsic charge symmetry breaking
will only become
transparent with more precise experimental data provided that concurrently a
proper accounting of Coulomb effects can be made.

For $\pi^-$ and $\pi^+$ elastic scattering from mirror nuclei (isospin
doublets) a more promising approach is the determination of ratios of
differential cross sections which can be combined into a super ratio.  For the
mirror nuclei $^{3}$H and $^{3}$He one can define two ratios
\begin{eqnarray}
R_1 = {d\sigma/d\Omega (\pi^{-} \, ^{3}H)\over d\sigma/d\Omega
(\pi^{+} \, ^{3}He)}
\,\,  ,
 \,\, R_2 = {d\sigma/d\Omega (\pi^{-} \,
^{3}He)\over d\sigma/d\Omega (\pi^{+} \, ^{3}H)}.
\end{eqnarray}

Clearly in the absence of Coulomb effects charge symmetry requires $R_1 = R_2
=1$.  The ratios $R_1$ and $R_2$ allow the determination of a super ratio
$R=R_1/R_2$, an experimental quantity independent of
calibrations of beam intensity and target thickness, which
can be arranged to be
also independent of detector inefficiencies.  Any deviation of $R$ from the
value 1.00 indicates \csb.  Significant deviations from 1.00 have been
measured\cite{BN90} which were explained in terms of direct and indirect
Coulomb  effects\cite{GG91}.  The latter include small differences in the
$^3$H and $^3$He wave functions:  differences in the odd nucleon radii and
the even nucleon radii for this pair of mirror nuclei. The numerical values are
in qualitative agreement with predictions obtained from solving the Faddeev
equations.

\vspace{0.5cm}
\noindent
{\em 3.5 Pion Reaction Experiments}

\vspace{0.2cm}
A comparison of the reactions $\pi^-d\rightarrow nn\pi^0$ and
$\pi^+d\rightarrow pp\pi^0$, as well as $\pi^-d\rightarrow nn\eta$ and $\pi^+d
\rightarrow pp\eta$ leads to the following ratios of the triple differential
cross sections

\begin{eqnarray}
R_1 = {d^3\sigma(\pi^-d \rightarrow nn\pi^0)\over d^3\sigma(\pi^+d
\rightarrow pp\pi^0)}
\end{eqnarray}
and
\begin{eqnarray}
R_2 = {d^3\sigma(\pi^-d \rightarrow nn\eta)\over d^3\sigma(\pi^+d \rightarrow
pp\eta)}.
\end{eqnarray}

Charge symmetry requires that both ratios $R_1$ and $R_2$ are equal to one for
all incident pion energies and for every angle of the scattered meson.
However, the $np$ mass difference and the Coulomb interaction will cause \csb.
Additional breaking will result from differences in the $n\eta$ and $p\eta$
interactions and from $\pi^0$-$\eta$ mixing.  An experiment to measure the
ratios $R_1$ and $R_2$ is in progress at the AGS of BNL\cite{NCP86}.  One of
the challenges for a successful completion of the experiment is to account
for any differences in the $\pi^-$ and $\pi^+$ incident beams.

\vspace{0.5cm}
\noindent
{\em 3.6 Breakup Reactions and Other}

\vspace{0.2cm}
\noindent
a) $^{4}He(\gamma,p)$ and $^{4}He(\gamma,n)$

\vspace{0.1cm}
Comparisons between the differential cross sections for the $^4$He$(\gamma,p)$
and $^4$He$(\gamma,n)$ reactions are a test of charge symmetry provided the
photon  absorption proceeds  purely by an E1 transition. Lately results have
been published on the absolute cross section of the $^{3}$He$(n,\gamma)$
reaction for five energies between 0.14 and 2.0~MeV to an accuracy of about
$\pm$10\%\cite{RK93}.  These results together with other $^{3}$He$(n,\gamma)$
capture data  applying detailed balance, and $^{4}$He$(\gamma,n)$
photo-disintegration data, if compared with $^{4}$He$(\gamma,p)$ data show that
the ratio of the photo-disintegration cross sections $R_\gamma =
\sigma[^{4}$He$(\gamma,p)]/\sigma[^{4}$He$(\gamma,n)]$ is equal to $\sim$1.1
over
the resonance region.  This ratio is also given by conventional theories which
leave out explicit charge symmetry breaking nucleon-nucleon  interactions.
Consequently the controversies stemming from the comparisons of the
$^{4}$He$(\gamma,n)$ and $^{4}$He$(\gamma,p)$ reactions may now have been
resolved.   Indeed, a new measurement of the $(\gamma,p)$ and $(\gamma,n)$
differential yields at 90$^\circ$ for the two-body photodisintegration of
$^4$He has now been reported\cite{REJF94}.   Using tagged photons of energies
between 25 and 60 MeV, data were obtained for both channels simultaneously
using windowless $\Delta E-E$ telescopes to detect the $^3$H and $^3$He recoil
particles.  The ratio of the angle integrated yields, which is insensitive to
many systematic errors due to the simultaneous measurement in the same
detectors, agrees with the results of calculations\cite{gcalcs} which take into
account only charge symmetric nucleon-nucleon interactions. Unfortunately, it
will be extremely difficult to perform the experiment and
carry out the analysis to a
precision sufficient for a quantitative deduction of charge symmetry breaking
effects.

Measurements of two-body photodisintegration of $^4$He at higher energies (in
the energy range 100 to 360 MeV), detecting the recoiling $^3$He, respectively
$^3$H nuclei, show ratios R within 30\% of unity at all angles with slightly
greater values at the higher photon energies.  In view of the errors of the
individual data points, the result is not inconsistent with a ratio of
unity\cite{RAS86}.

\vspace{0.5cm}
\noindent
b) $^{2}H(\vec{d},pn)d$ and $^{2}H(\vec{d},np)d$

\vspace{0.1cm}
In a recent  publication\cite{CH93} a novel probe of charge symmetry breaking
was reported involving single deuteron breakup in polarized $dd$ scattering:
$\vec{d}d\rightarrow dpn$.  In the first phase of the experiment the spin
observables $A_y, A_{yy}$ and $A_{zz}$ are compared for the mirror reactions
$^2$H$(\vec{d},dp)n$ and $^2$H$(\vec{d},dn)p$ at two angle pairs:
$(\theta_d,\phi_d,\theta_N,\phi_N)$ = $(17.0^\circ, 0^\circ, 17.0^\circ,
180^\circ)$ and $(17.0^\circ, 0^\circ, 34.5^\circ, 180^\circ)$.  The incident
energy is rather low so the unwanted  effects of the Coulomb interaction are
expected to be large.

Thus the comparison of $dp$ and $dn$ coincidence
data under identical kinematical
conditions serves to check the influence of the Coulomb interaction.  Only for
the angle pair $(17.0^\circ, 0^\circ, 34.5^\circ, 180^\circ)$ is there a
statistically significant difference in the measured values for $A_{zz}$
(0.060$\pm$0.025 between the dp and dn $A_{zz}$ data).  In the second phase of
the experiment the reaction $^2$H$(\vec{d}, pn)d$ was measured at
$(\theta_p,\phi_p,\theta_n,\phi_n)$ = $(17.0^\circ, 0^\circ, 17.0^\circ,
180^\circ)$.  In this case if charge symmetry holds then $A_{yy}$ and $A_{zz}$
along the kinematic locus in a plot of  $T_p$ versus $T_n$ should be symmetric
with respect to the  $T_p = T_n$ point and $A_y$ should be antisymmetric with
respect to this equal energy point.  The result indicates that $A_{yy}$ and
$A_{zz}$ are indeed symmetric to within their statistical uncertainties of
$\pm$0.025, while $A_y$ is antisymmetric to within their statistical
uncertainties of $\pm$0.016.  Only when taking restricted energy intervals may
there be an indication of \csb.  Clearly this new method has %%
promise provided the incident energy is increased, in order  to be more certain
about the effects of the Coulomb interaction, together with an improved
statistical accuracy.

\vspace{0.5cm}
\noindent
{\bf 4. Summary and Suggested Further Work}

\vspace{0.4cm}

Charge independence and charge  symmetry breaking is caused by the $d$-$u$
quark mass difference $m_d-m_u >0$, along with electromagnetic effects. The
consequences of these effects can be manifest by including their  imprint on
hadronic matrix elements or by  using quark models directly. Lattice QCD
calculations are not yet  sufficiently accurate to handle small differences
between the small masses  of current quarks.

The general goal of this area of research is to find small charge  independence
and charge symmetry breaking effects, and then explain these in terms  of
fundamental ideas. Over the years  there has indeed been substantial
experimental and theoretical progress,  which we now summarize briefly. First,
we recall the idea that $m_d-m_u>0$ accounts for  the observed mass
differences between members of hadronic isospin  multiplets. Next we note that
charge independence breaking in the $^1$S$_0$ system is well explained   in
terms of meson exchange models. The most significant effect is the 4.6~MeV
mass difference between charged and neutral pions, which has its ultimate
origin in the Coulomb interaction amongst the quarks. The effects of
$\gamma\pi$ exchange are also relevant.

Substantial effects of $\rho^0$-$\omega$ mixing have been observed in the
$e^+e^-
\rightarrow \pi^+\pi^-$ cross section at $q^2\approx m^2_\omega$. The results
allow an extraction of the strong contribution to the $\rho^0$-$\omega$ mixing
matrix element $<\rho^0|H_{str}|\omega>\approx$ - 5200~MeV$^2$. Two nucleons
may exchange a mixed $\rho^0$-$\omega$ meson. If one uses
$<\rho^0|H_{str}|\omega>\approx$ - 5200~MeV$^2$ one obtains a nucleon-nucleon
interaction which accounts for $\Delta a_{CSB} = a^N_{pp}-a^N_{nn} = 1.5\pm
0.5$ fm. Such a force also is responsible for most of the strong interaction
contribution to the $^3$H-$^3$He binding energy difference and accounts for
much of the Nolen-Schiffer anomaly. More generally, the use of potentials
consistent with $\Delta a_{CSB}$ and $\Delta a_{CD}$ accounts for formerly
anomalous binding energy differences in mirror nuclei and in analog states.

The TRIUMF (350 and 477~MeV) and IUCF (183~MeV) experiments  have compared
analyzing powers of $\vec np$ and $n\vec p$ scattering and observe charge
symmetry breaking at the
level expected from $\pi,\gamma$ and $\rho^0$-$\omega$ exchange effects. The
latter effects are important at 183~MeV.

The virtual mesons of nucleon-nucleon potentials have space-like  four-momenta,
while the $\rho^0$-$\omega$ mixing matrix element has been extracted for
on-shell vector  mesons. Several authors have  postulated significant off-shell
effects  that essentially wipe out the effects of $\rho^0$-$\omega$ mixing in
nucleon-nucleon interactions. However, these ideas seem to contradict  the
feature that the $\gamma\rho$ or $\gamma^*\rho$
transition matrix element is observed to be
approximately independent of the photon four momentum.

Charge symmetry breaking has also been studied in hypernuclear systems
and the spin dependence of the $\Lambda N$ interaction has been extracted
from the observed masses of the ground and first excited state.

\vspace{0.5cm}
 \noindent
{\em 4.1 Suggested Experimental Investigations}

\vspace{0.2cm}
Experimental searches for charge independence and charge symmetry breaking are
difficult, time consuming tasks.  So the first order of business is to
encourage the  ongoing experiments. Thus we look forward to the complete
analysis of the TRIUMF comparison of neutron and proton analyzing powers in
$np$ elastic  scattering at 350 MeV.
The reaction $\pi^- d\rightarrow \gamma nn$ is the
principle  source of information on low energy nn scattering. Thus we
anticipate the
the analysis of the data taken in the new TRIUMF measurement. The
asymmetry about 90$^\circ$  in the angular  distribution of  the $np
\rightarrow d \pi^0$ reaction  would provide information about $\pi^0$-$\eta$
mixing and we encourage a precision experiment. A BNL comparison of the
reactions $\pi^-d\rightarrow nn\pi^0$ and $\pi^+d\rightarrow pp\pi^0$, as well
as $\pi^-d\rightarrow nn\eta$ and $\pi^+d \rightarrow pp\eta$ would allow a new
study of the difference between the nn and pp  interactions as well as
$\pi^0$-$\eta$ mixing. We also encourage the extension of the
$^{2}$H$(\vec{d},pn)d$ and $^{2}$H$(\vec{d},np)d$ measurements to higher
energies.

Next we discuss possibilities for improving earlier experiments and some  new
ones. A clear observation of a non-zero value of the
$dd \rightarrow$$^4$He$\pi^0$ cross section would provide a definite
manifestation of charge
symmetry breaking. One needs to learn how to  handle the background from the
$dd \rightarrow$$^4$He$\gamma$ reaction. More precise data for $\pi^\pm$ -
$^3$He ($^3$H) elastic scattering would  allow a more precise determination of
the difference between the even and  odd radii.  The masses of the hypernuclei
should be re-examined and determined more  precisely.  We have been mainly
concerned with nuclear effects, but also stress that  the effects
of $\Lambda$-$\Sigma^0$ mixing could be observed by  comparing the weak decay
rates  for $\Sigma^+\rightarrow\Lambda e^+\nu_e$ and
$\Sigma^-\rightarrow\Lambda e^-\bar{\nu}_e$.

\vspace{0.5cm}
\noindent
{\em 4.2 Suggested Theoretical Explorations}

\vspace{0.2cm}

There are a number of tasks which can be readily accomplished with  existing
techniques. The $\gamma \pi$ exchange potential should be  re--evaluated,
especially its consequences for the spin--dependent neutron--proton scattering.
The role of vector meson mixing for charge symmetry breaking $\Lambda N$
scattering should be re-examined. Careful evaluations of the effects of the
different densities seen by valence neutrons and protons should be  included in
QCD sum rule evaluations of the medium modifications to the  neutron-proton
mass difference.
Realistic charge dependent and  charge
asymmetric potentials should be used in shell model calculations; this
awaits a more complete evaluation of the $\gamma \pi$ exchange potential.
The scattering theory for pion interactions with $^3$He
($^3$H) could be made  more precise so as to better learn about charge symmetry
violations in the  wave functions. The role of Coulomb effects in the $\pi$
deuteron total  cross section should be re-assessed to confirm the conclusions
of the  original experiment.
The effects of charge symmetry breaking in the $\pi^0$-nucleon interaction
on the $np\rightarrow d \pi^0$ angular distribution  should be estimated.
Similarly one should make careful predictions for the
$\pi d \to NN\pi^0$ and $\pi d \to NN\eta$ reactions.

There are also some more challenging tasks. One is to find a precise method  to
determine the charge dependence of meson-nucleon coupling constants. Present
data limits such effects to less than about 1\%, which would allow significant
effects. The off-shell dependence of $\rho^0$-$\omega$ mixing  should be
determined in a definitive way. We have argued here that  a significant q$^2$
variation contradicts many experiments involving $\gamma\rho$ transitions for
real and virtual photons. But it would be interesting to see if  a model with
significant off-shell dependence in the $\rho^0$-$\omega$ mixing could be
consistent with those experiments.  We have discussed above explanations  of
the Nolen--Schiffer  anomaly which involve  both vector and scalar effects. The
vector explanations seem closely  related to two-nucleon data  and therefore
more likely correct. However, scalar effects are  surely present. It would be
useful to devise theoretical and experimental  means to separate and
distinguish the two categories of explanation. In either case, the ultimate
origin of the anomaly is the mass difference  between up and down quarks.

This  quark mass difference, $m_d-m_u$ seems to be related to a large variety
of phenomena in particle and nuclear physics.  Most of the effects are well
understood.  Perhaps the next relevant question is why are there two light
quarks with a slightly different mass?

\vspace{0.2in}
\noindent
{\bf 5. Acknowledgments}

\vspace{0.5cm}
This work was supported in part by the USDOE and by the Natural Sciences
and Engineering Research Council of Canada.

\vspace{0.5cm}

\noindent {\bf 6. References}

\end{document}